\documentclass[pdflatex,sn-nature]{sn-jnl}%

\usepackage{graphicx}%
\usepackage{multirow}%
\usepackage{amsmath,amssymb,amsfonts}%
\usepackage{amsthm}%
\usepackage{mathrsfs}%
\usepackage[title]{appendix}%
\usepackage{xcolor}%
\usepackage{textcomp}%
\usepackage{manyfoot}%
\usepackage{booktabs}%
\usepackage{algorithm}%
\usepackage{algorithmicx}%
\usepackage{algpseudocode}%
\usepackage{listings}%
\usepackage{bbm}

\usepackage{siunitx}%
\usepackage{subcaption}%
\usepackage{float}%

\graphicspath{{figures/}}

\theoremstyle{thmstyleone}%
\theoremstyle{thmstyletwo}%
\theoremstyle{thmstylethree}%

\begin{document}

\title{Skillful Thunderstorm Forecasts Learned Directly from
Observations}

\author[1]{\fnm{Jaideep} \sur{Pathak}}
  \equalcont{Jaideep Pathak, Mohammad Shoaib Abbas, and Peter Harrington contributed equally to this work.
  \par\textsuperscript{$\ddagger$}Dale Durran and Mike Pritchard jointly advised this work.}

  \author[1]{\fnm{Mohammad Shoaib} \sur{Abbas}}
  \equalcont{Jaideep Pathak, Mohammad Shoaib Abbas, and Peter Harrington contributed equally to this work.
  \par\textsuperscript{$\ddagger$}Dale Durran and Mike Pritchard jointly advised this work.}

  \author[1]{\fnm{Peter} \sur{Harrington}}
  \equalcont{Jaideep Pathak, Mohammad Shoaib Abbas, and Peter Harrington contributed equally to this work.
  \par\textsuperscript{$\ddagger$}Dale Durran and Mike Pritchard jointly advised this work.}

  \author[2]{\fnm{Corey} \sur{Potvin}}
  \author[2]{\fnm{Adam} \sur{Clark}}
  \author[3,2]{\fnm{Patrick} \sur{Skinner}}
  \author[1]{\fnm{Aayush} \sur{Gupta}}
  \author[1]{\fnm{Tao} \sur{Ge}}
  \author[1]{\fnm{Suman} \sur{Ravuri}}
  \author[1]{\fnm{Corey} \sur{Adams}}
  \author[1]{\fnm{Noah} \sur{Brenowitz}}
  \author[1]{\fnm{Alberto} \sur{Carpentieri}}
  \author[1]{\fnm{Jason} \sur{Stock}}
  \author[1]{\fnm{Jussi} \sur{Leinonen}}
  \author[1]{\fnm{Zeyuan} \sur{Hu}}
  \author[1]{\fnm{Oliver} \sur{Hennigh}}
  \author[1]{\fnm{Nicholas} \sur{Geneva}}
  \author[1]{\fnm{Mike} \sur{Pritchard}\textsuperscript{$\ddagger$}}
  \author[1]{\fnm{Dale} \sur{Durran}\textsuperscript{$\ddagger$}}
  \affil[1]{\orgname{NVIDIA Corporation},
  \orgaddress{\city{Santa Clara}, \state{California}, \country{USA}}}

  \affil[2]{\orgname{National Severe Storms Laboratory (NSSL)},
  \orgaddress{\city{Norman}, \state{Oklahoma}, \country{USA}}}
  \affil[3]{\orgname{Cooperative Institute for Severe and High-Impact Weather Research and Operations (CIWRO)},
  \orgaddress{\city{Norman}, \state{Oklahoma}, \country{USA}}}

\abstract{Nowcasting the evolution of severe convective storms over the first few hours after observations are collected is critical for issuing timely, accurate warnings. Forecast accuracy is hindered because the variables that matter most, such as rain, hail, and lightning, are imperfectly represented by numerical model physics and are difficult to initialize correctly. Here we introduce StormScope, a computationally efficient generative forecasting system that learns storm-scale evolution directly from satellite, radar, and lightning observations over the continental United States with no numerical model used for training or initialization. StormScope produces coherent, physically plausible storm trajectories with well-calibrated uncertainty. Across more than 900 forecasts, it greatly exceeds the National Oceanic and Atmospheric Administration's state-of-the-art forthcoming operational numerical modeling system in both deterministic and probabilistic skill through the 0--2-h nowcasting window. StormScope's skill for strong radar echoes exceeds the operational state of the art by approximately 75\% and 45\% at lead times of 30~min and 2~h, a relative improvement several times larger than machine learning has delivered for global medium-range forecasting. Crucially, it also anticipates storms that develop after forecast initialization, establishing its utility beyond extrapolation of existing radar echoes. It likewise surpasses a leading AI nowcasting model for intense precipitation. Together, these results establish learning directly from observations as a route toward skillful, well-calibrated storm-scale forecasts that is unconstrained by the initialization and microphysics bottlenecks limiting numerical models and their emulators.}


\maketitle

\section*{Main}

Accurate forecasting of damaging thunderstorms and convective systems is critical for severe-weather warnings, flash-flood guidance, aviation safety, and emergency response. Yet numerical forecasting of such storms is more difficult than larger-scale global numerical weather prediction (NWP). Rain, hail, and lightning are not predicted from first principles but diagnosed through approximate microphysical parameterizations. Individual precipitating cells within a thunderstorm can develop or dissipate in just 15 minutes. Even a 2-h forecast must correctly follow the evolution of vigorous convective updrafts over multiple cycles of growth and decay. Small errors in the specification of the atmospheric state amplify during these cycles, limiting the predictability of the detailed placement of thunderstorm precipitation to a few hours \citep{Surcel2015-qk}.

Appropriate initial conditions are difficult to specify at the kilometer scale. Key forecast variables in kilometer-scale models involve difficult-to-observe moisture fields. The variety of sizes and types of hydrometeors suspended in a thunderstorm is too large and complex for their complete initial distribution to be determined. State-of-the-art kilometer-scale NWP models, such as the High-Resolution Rapid Refresh (HRRR)~\citep{dowell2022high} and the Warn-on-Forecast System (WoFS)~\citep{heinselmanWarnonForecast2024}, greatly simplify cloud microphysics through parameterizations, collapsing this complexity into far fewer fields. Yet even these parameterized fields are difficult to determine from observations. Storms in kilometer-scale models typically require tens of minutes to spin up, substantially limiting forecast skill at crucial warning lead times. WoFS requires more than 15 min for data assimilation and forecast production. This delay reduces the value of storm-scale NWP to operational meteorologists issuing timely warnings. 

\begin{figure}[H]
    \centering
    \includegraphics[width=\linewidth]{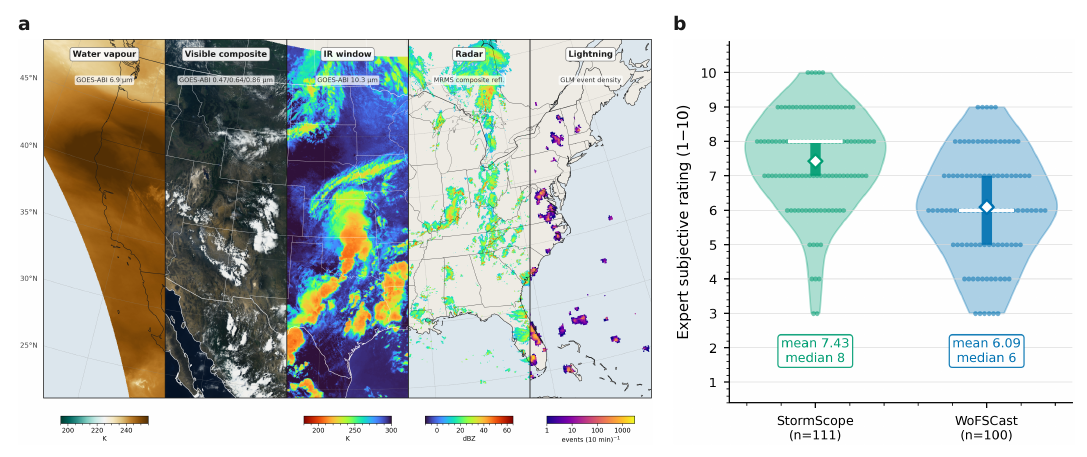}
    \caption{\textbf{Illustration of StormScope's multimodal forecast state over the continental United States (CONUS) domain and subjective evaluation scores.} \textbf{a}, Selected satellite channels ingested by the model at 20:50~UTC on 12 July 2025 shown as vertical slices on the model's native CONUS grid, from west to east: mid-level water-vapor brightness temperature (6.9~\textmu m); a composite of blue, red, and near-infrared channels (0.47, 0.64, and 0.86~\textmu m); clean infrared window (10.3~\textmu m); Multi-Radar/Multi-Sensor (MRMS) composite radar reflectivity; and GLM lightning-event density. \textbf{b}, Subjective ratings of StormScope and WoFSCast by operational meteorologists during the NOAA-NSSL 2026 Hazardous Weather Testbed Spring Forecasting Experiment.}
    \label{fig:intro_combined}
\end{figure}

Here we present StormScope, a generative machine learning (ML) model trained on radar, lightning, and multispectral satellite observations. StormScope's input data, illustrated in Fig.~\ref{fig:intro_combined}a, are collected from eight visible and infrared channels of the Advanced Baseline Imager (ABI) and the Geostationary Lightning Mapper (GLM) on the NOAA GOES satellites and combined with NOAA's Multi-Radar/Multi-Sensor (MRMS) radar mosaic over the continental United States. StormScope autoregressively updates these fields every 10 minutes, producing both deterministic and ensemble forecasts with continental coverage.

Previous ML methods for kilometer-scale convective modeling train on combinations of observational data and convection-allowing models like HRRR or WoFS \citep{pulkkinen2019pysteps,ravuri2021skilful, leinonen2023latent,gong2024cascast, dai2024four,chase2025score,deluca2025nowcasting}. One such approach, NowcastNet \citep{zhang2023skilful}, outperformed earlier ML-based and numerical extrapolation schemes for convective forecasting, including pySTEPS \citep{pulkkinen2019pysteps, bowler2006steps} and DGMR \citep{ravuri2021skilful}.  

Below, we show that StormScope outperforms leading NWP (WoFS) and ML models over a wide range of deterministic and probabilistic metrics, particularly at the highest intensities. Case studies of extreme events also reveal visually striking improvements in the location, intensity, and structure of forecast storms. StormScope's superior performance, combined with its computational efficiency and its ability to be 
initialized rapidly from observations, offers great potential for improving severe-weather warnings.

\section*{Comparison of forecast storm structure and subjective forecaster evaluation}\label{sec:qualitative_skill}

Operational meteorologists' subjective ratings during the 5-week Hazardous Weather Testbed (HWT) Spring Forecasting Experiment (SFE)~\cite{Clark2026SFE} corroborate StormScope's physical realism and operational utility. StormScope forecasts received a median subjective quality score of 8/10 across 111 evaluations (Fig.~\ref{fig:intro_combined}b). Without a direct forecaster comparison of WoFS in this testbed, we benchmark against WoFSCast, a kilometer-scale ML emulator of WoFS, which closely matches WoFS across quantitative and qualitative measures~\cite{flora2025wofscast}. Forecasters clearly preferred StormScope over WoFSCast.

\begin{figure}
    \centering
    \includegraphics[width=0.8\linewidth,height=0.64\textheight,keepaspectratio]{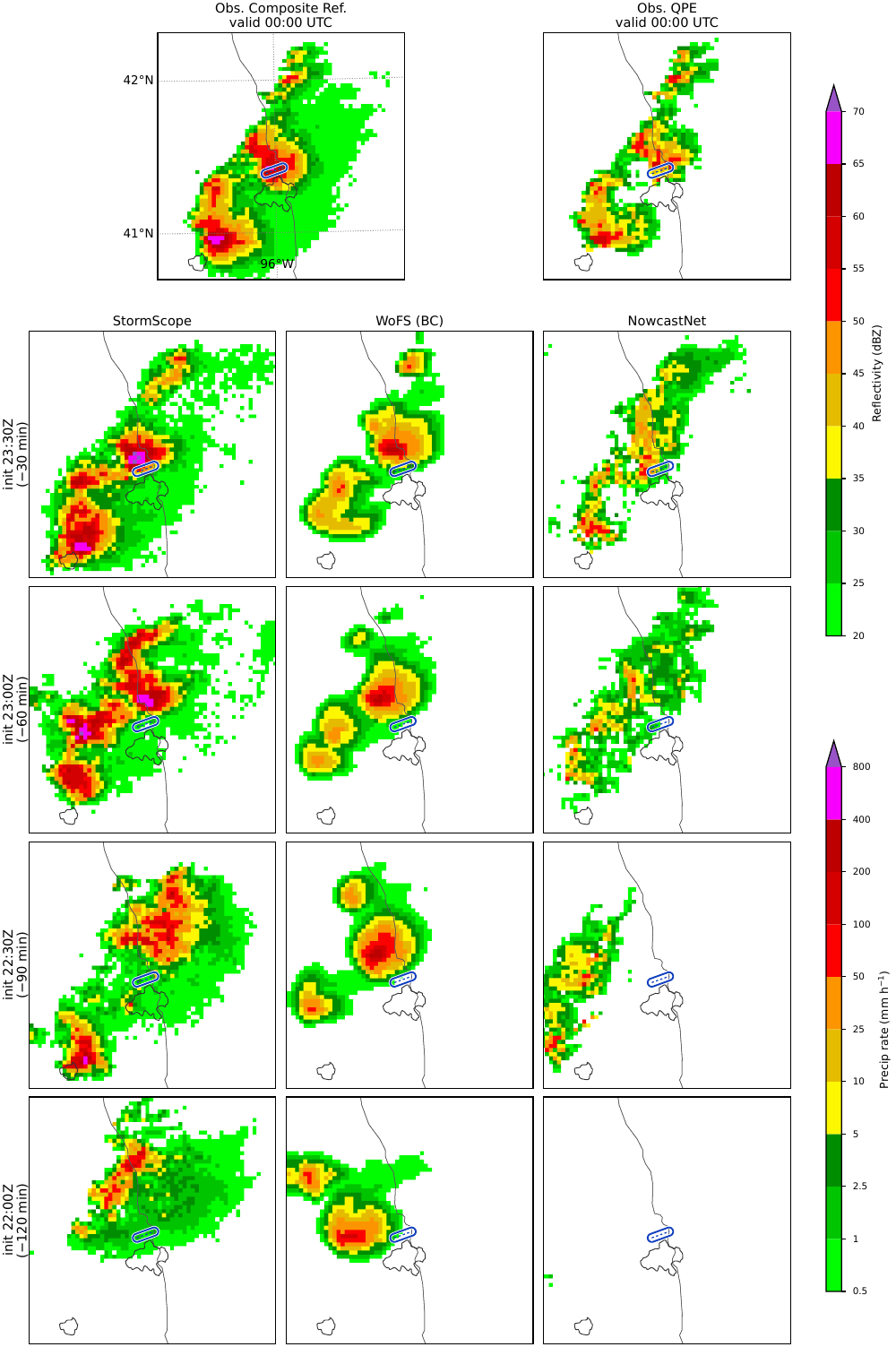}
    \caption{\textbf{Forecasts of a tornadic thunderstorm near the Nebraska/Iowa border.}  Top row: observed MRMS composite reflectivity (left) and corresponding quantitative precipitation estimate (QPE, right) at 00:00 UTC on 18 April 2025. Rows 2--5 show forecasts initialized at lead times of 30, 60, 90 and 120 min. Left, middle, and right columns show forecasts from StormScope, WoFS, and NowcastNet, respectively. The StormScope and WoFS (NowcastNet) forecasts should be verified against the MRMS composite reflectivity (QPE analysis). The blue curve outlines the track of the tornado occurring at the verification time.
    }
    \label{fig:ne-ia-case}
\end{figure}

As an extreme-weather example, consider forecasts of two supercells that formed near the Nebraska-Iowa border on 17 April 2025, producing  baseball-sized hail and 3 confirmed tornadoes between 23:55 and 00:15 UTC, including an EF-3 tornado from the northern cell \cite{NWSOmaha2025April17}. Figure~\ref{fig:ne-ia-case} compares progressively longer-lead forecasts from StormScope, WoFS and NowcastNet, all verifying at 00:00 UTC on 18 April 2025, during the EF-3 tornado. All forecasts depict results from the first indexed ensemble member. Forecasts initialized 2 h before verification show both WoFS and StormScope depicting an intense convective event in approximately the correct location, whereas NowcastNet misses the signal almost entirely. In forecasts issued 90 and 60 min before verification, StormScope suggests two or three very intense storms in roughly the correct location with radar reflectivities reaching observed values of 65--70 dBZ. NowcastNet's tendency to forecast disorganized distributions of high-intensity precipitation that do not reflect the underlying storm structure also appears in the additional severe-weather case studies in Figs.~\ref{fig:stl}--\ref{fig:line}.  At 30-min lead time, only the StormScope forecast shows intensity patterns suggestive of the supercell thunderstorm that spawned the tornado.

\section*{Probabilistic and deterministic forecast skill}\label{sec:quantitative_skill}

\subsection*{Skill relative to a leading operational model}

WoFS is our baseline because it is a key NWP tool supporting National Weather Service (NWS) operations in the watch-to-warning time frame, providing rapidly updating probabilistic guidance at 0–3-h lead times. We compare StormScope with all 929 quality-controlled WoFS forecasts from the 2025 severe-weather season (March--August), spanning roughly 60 calendar days selected by meteorologists for their severe-weather potential. Matching domains and verification times, we evaluate deterministic skill using three metrics: the Critical Success Index (CSI), Probability of Detection (POD), and Success Ratio (SR), computed for 40- and 55-dBZ radar-echo thresholds within all 33 $\times$~33~km neighborhoods. Sixteen-member ensemble forecasts are evaluated over the same neighborhoods with the probabilistic Fractions Skill Score (pFSS)~\cite{necker} and ensemble reliability diagrams \cite{murphy-reliability}. Further details are provided in Methods. WoFS reflectivity is bias-corrected (Extended Data Sec.~\ref{secA:wofs_bias_correction}) to improve its performance. StormScope requires no bias correction.

StormScope's POD for at least 40-dBZ echoes (moderate to heavy rain~\cite{NOAA_VIP}) shows that it correctly catches 75\%, 64\%, and 50\% of the echoes at lead times of 30, 60, and 120~min, respectively, compared with 47\%, 40\%, and 34\% for bias-corrected WoFS (Extended Data Table~\ref{tab:wofs_vs_stormscope_skill}). These roughly 50\% improvements in POD are achieved while  reducing false alarms; StormScope's SR values are 75\%, 65\%, and 53\% compared to 55\%, 48\%, and 45\% for WoFS.

Although absolute performance is lower than at 40~dBZ, the results for more extreme 55-dBZ echoes (very heavy rain or hail) show even greater relative superiority for StormScope through 1~h. StormScope's POD and SR for these intense storms are roughly double those of WoFS. At 2 h, WoFS has slightly higher POD than StormScope, but the chance of correctly capturing observed 55-dBZ echoes is about 10\% or less for both models.

\begin{figure}[H]
    \centering
    \includegraphics[width=\linewidth]{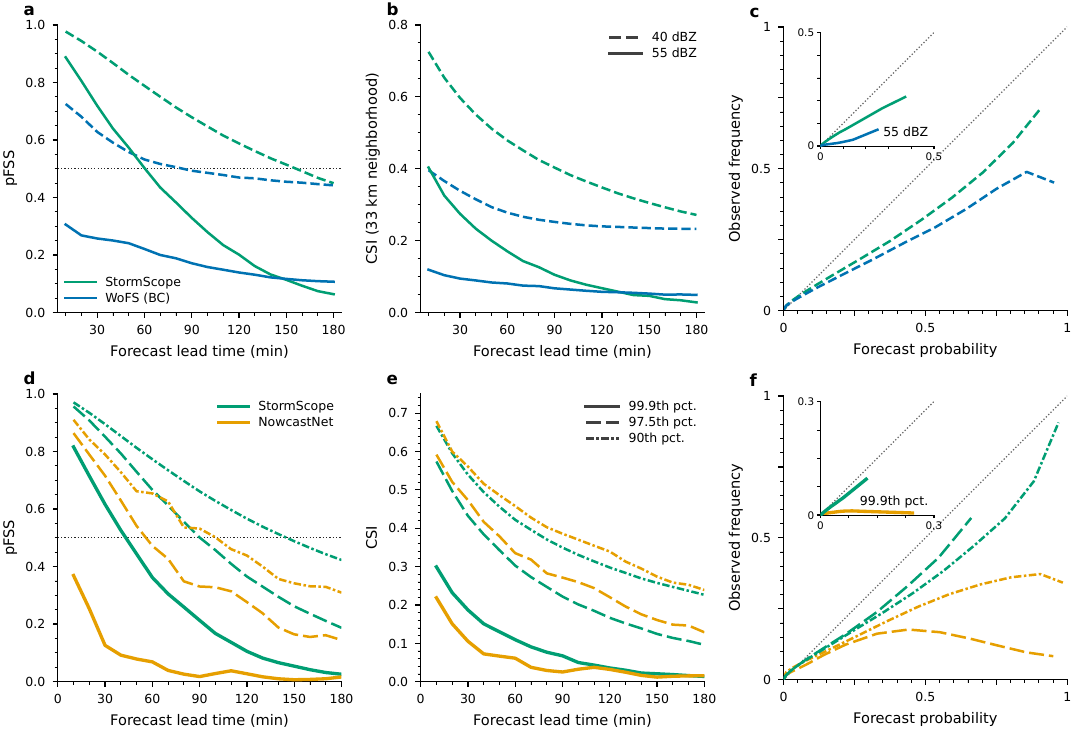}
    \caption{{\bf Quantitative skill scores comparing StormScope with WoFS and NowcastNet aggregated across 929 forecasts.}  StormScope, WoFS and NowcastNet are shown in green, blue, and orange, respectively. Dashed and solid lines denote 40 and 55~dBZ thresholds in the StormScope--WoFS comparison \textbf{a--c}. Dash-dot, dashed, and solid lines denote the 90th, 97.5th, and 99.9th percentiles in the StormScope--NowcastNet comparison \textbf{d--f}.  Probabilistic Fractions Skill Score (pFSS) \textbf{a,d}; neighborhood Critical Success Index (CSI) \textbf{b,e}; and ensemble reliability diagrams \textbf{c,f}.  Horizontal dotted lines in \textbf{a} and \textbf{d} indicate the cutoff for skillful pFSS forecasts. Gray 1:1 lines in \textbf{c} and \textbf{f} are the references for perfect reliability.
    }
    \label{fig:combined_skill_figure}
\end{figure}

CSI (Eq.~\ref{eqn:contingency}) penalizes forecasts for missed events and false alarms. At the 40-dBZ threshold, CSI values for StormScope remain superior to WoFS through 180~min (Fig.~\ref{fig:combined_skill_figure}b). Although absolute performance is again lower than at 40~dBZ, the result for more extreme 55-dBZ echoes shows even greater superiority for StormScope, with improvements by factors of 2.9 at 30~min and 2.1 at 60~min. Little skill remains for either model at 3~h, although the WoFS score is slightly higher. Object-based CSI evaluations in Extended Data Sec.~\ref{secA:objcomp} show score improvements of a similar magnitude.

Figure~\ref{fig:combined_skill_figure}a shows StormScope's pFSS exceeds that of WoFS across all lead times at the 40-dBZ threshold. Noting that a pFSS value of 0.5 is the nominal threshold for skillful forecasts of phenomena that cover a small fraction of the neighborhood~\citep{roberts2008scale}, we find that StormScope retains useful  skill out to 160~min compared with 80~min for WoFS. At the 55-dBZ level, StormScope's pFSS remains above the useful-skill threshold out to about 60~min, whereas WoFS scores fall below it at all lead times. StormScope also outperforms Eulerian and Lagrangian extrapolation baselines at all lead times and thresholds (Extended Data Sec.~\ref{secA:extrapolation}).

Reliability diagrams (Fig.~\ref{fig:combined_skill_figure}c) compare the observed frequency of an event against the predicted probability of the event within the ensemble. For 40- and 55-dBZ threshold echoes aggregated over 0--3~h forecasts, StormScope is more reliable at all probabilities, particularly at 55~dBZ, although both models overestimate event probabilities. The curves for the 55-dBZ threshold terminate at forecast probabilities below 0.4 because stronger echoes are seldom forecast by a majority of the ensemble members.

\subsection*{Skill relative to a leading AI model}\label{sec:rainfall}

We compare StormScope with NowcastNet~\cite{zhang2023skilful}, which uses sequences of radar-derived precipitation fields to generate ensemble forecasts up to 3~h ahead. It combines an evolution network that learns precipitation motion and intensity changes with a conditional generative adversarial network that supplies finer-scale precipitation structure. We chose NowcastNet because it has outperformed several other ML-based nowcasting schemes.

Using the previously considered 929 forecast initializations, we evaluate forecasts at 3-km resolution on domains centered on and closely approximating the WoFS domains. We generate NowcastNet forecasts on its native MRMS $1024\times 1024$-pixel, $0.01^{\circ}$-latitude-longitude grid using the authors' prescribed protocol and code.
Because the two systems are natively verified against different fields (StormScope against MRMS base reflectivity and NowcastNet against MRMS quantitative precipitation estimates), we score both models on rainy-population percentile thresholds computed per model against its corresponding verification field, placing them in a common framework (Extended Data Sec.~\ref{secA:NowcastNet}). 

Figure~\ref{fig:combined_skill_figure}d--f compares skill at the 90th, 97.5th, and 99.9th percentiles, corresponding to 35, 45, and 55~dBZ, respectively, roughly corresponding to moderate rain, heavy rain, and very heavy rain or hail~\cite{NOAA_VIP}.  For deterministic forecasts, StormScope has higher CSI for very heavy rain or hail (99.9th percentile) over the first 100~min, while NowcastNet performs slightly better for light to moderate rain (90th percentile) and heavy rain (97.5th percentile) at all lead times. Neighborhood CSI, again computed over $33\times33$~km neighborhoods, does not fully capture differences in storm structure. In these extreme-event case studies, NowcastNet produces fragmented precipitation patterns that poorly reproduce the observed organization, whereas StormScope captures this organization more faithfully (Fig.~\ref{fig:ne-ia-case}; Figs.~\ref{fig:stl}--\ref{fig:line}).

Probabilistic metrics show larger differences in ensemble forecast performance. StormScope's pFSS exceeds that of NowcastNet at every threshold and lead time (Fig.~\ref{fig:combined_skill_figure}d). StormScope's pFSS at the 99.9th-percentile intensity stays above the 0.5 threshold for skillful forecasts for almost 45~min, whereas NowcastNet never exceeds it, even at 10~min. In the reliability diagram (Fig.~\ref{fig:combined_skill_figure}f), StormScope's ensemble forecasts provide more reliable guidance at all probabilities and all three percentile thresholds than those from NowcastNet. 

At convective nowcasting scales, forecasts have a high degree of uncertainty. Models designed for operations such as WoFS must provide ensemble forecasts to characterize this uncertainty. StormScope, as part of its design, provides far better probabilistic guidance than NowcastNet through significantly improved ensemble calibration. 

\subsection*{Forecasting convective initiation and intensification}\label{sec:ci}

A key question in nowcasting is whether a system can forecast the formation of intense convective storms that are much weaker or absent at initialization. We define ``storm age'' as time of first echo observation relative to forecast initialization. The age is positive if observed before initialization and negative otherwise. We compare StormScope and WoFS using object-based verification stratified by storm age~\citep{guerra2022quantification, skinner_2025}, matching storms that include at least one 45-dBZ echo within overlapping 40-dBZ regions (Methods Sec.~\ref{sec:object-based-eval}).

\begin{figure}[H]
    \centering
    \includegraphics[width=\linewidth]{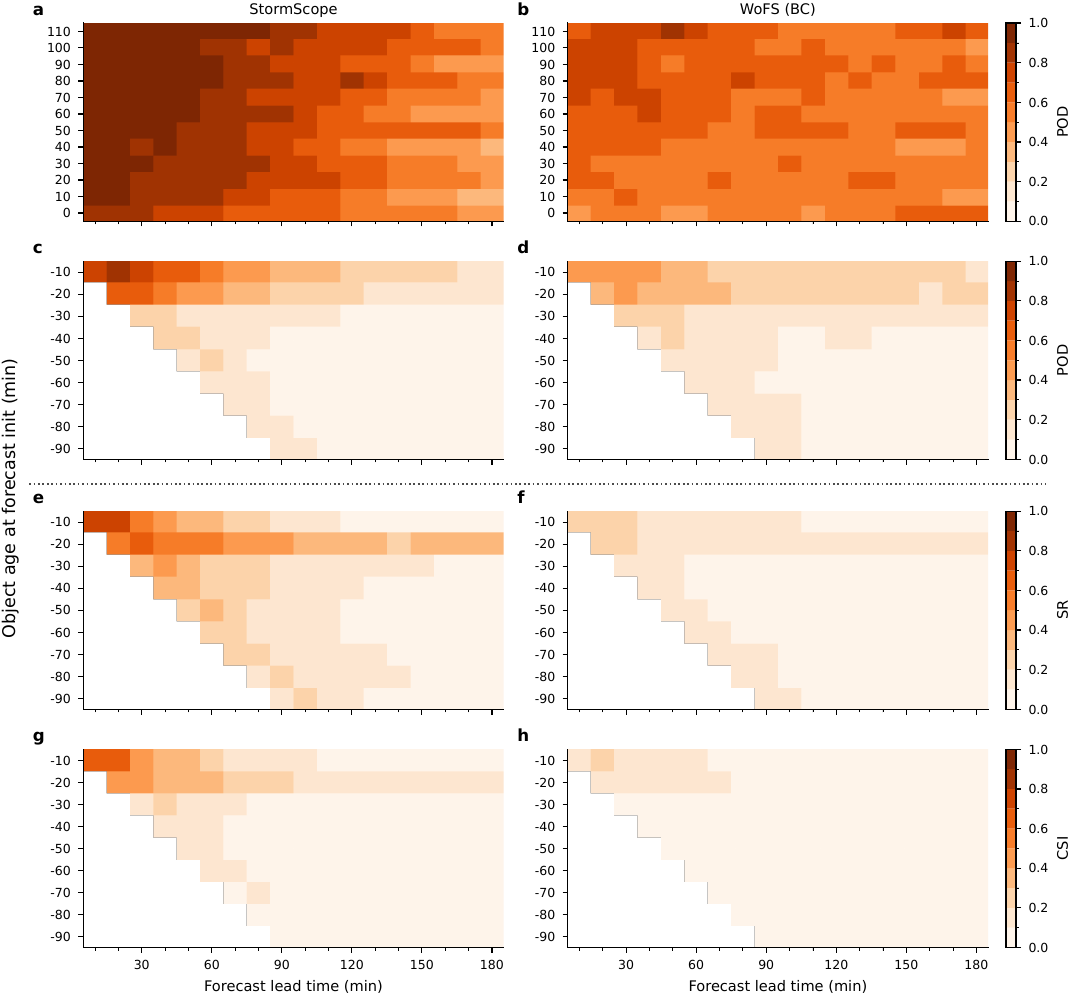}
    \caption{\textbf{Object-matching-based skill metrics for StormScope and WoFS as a function of storm age and forecast lead time.} Probability of detection (POD) for storms with radar echoes exceeding 45~dBZ that exist when forecasts are initialized, for \textbf{a}, StormScope and \textbf{b}, WoFS, binned by positive age. Panels \textbf{c} and \textbf{d} are as in \textbf{a} and \textbf{b}, except for storms that develop after initialization and therefore have negative age (see text). Panels \textbf{e} and \textbf{f} show Success Ratio (SR), and panels \textbf{g} and \textbf{h} show Critical Success Index (CSI), for the same negative-age storms.}
    \label{fig:convective_initiation_wofs_ss}
\end{figure}

For storms already present at initialization, StormScope has higher POD across age bins through the first 90~min (Fig.~\ref{fig:convective_initiation_wofs_ss}a,b). For storms aged at least 60~min, StormScope detects at least 90\% of these storms through forecast lead times of 50~min, whereas WoFS never achieves a POD of 90\%. The advantage is narrower at longer lead times.

Detection generally improves with storm age in both models. For storms aged 110~min, StormScope retains POD$\geq 0.7$ through 140~min, outperforming WoFS (Fig.~\ref{fig:convective_initiation_wofs_ss}a, b) and illustrating the advantage in detection horizon for older storms. Previous research suggests that repeated data-assimilation cycles contribute to this age dependence in WoFS~\cite{guerra2022quantification}. StormScope's input window is fixed at one hour, so its continued improvement between ages 70 and 110~min cannot be explained simply by a longer observation history. It may instead reflect older storms belonging to well-organized, long-lived convective systems with more predictable evolution.

For convective initiation and intensification, we continue to define the age of observed storms by the now-negative difference between the time they first appear and the time the forecast is initialized. The age of storms that develop in the forecast is defined similarly using their first appearance in that forecast. Object matching includes spatial matches and temporal matches within $\pm 10$~min. Forecast storms appearing in the first time step (age $=-10$~min) are unmatched if their only possible match is a storm present at forecast initialization. POD, SR, and CSI are plotted as functions of forecast lead time and storm age (Fig.~\ref{fig:convective_initiation_wofs_ss}c--h), with hits and misses binned by observed storm age and false alarms by forecast storm age (Methods Sec.~\ref{sec:object-based-eval}).

StormScope substantially outperforms WoFS in all three metrics on storms that develop 45-dBZ echoes within 20~min of forecast initialization, although both models' skill clearly declines at longer lead times. StormScope's skill in convective-initiation and intensification forecasts at short lead times is likely due in part to its ingestion of satellite imagery showing growing cumulus congestus clouds and radar reflectivity showing initial weak echo development.  Nevertheless, as its low false alarm ratio (high SR) indicates, StormScope effectively identifies clouds that grow into cumulonimbus.  

\section*{Conclusions}

Together, these results establish that storm-scale evolution can be learned directly from observations and that StormScope surpasses the operational numerical state of the art through the crucial first hours of the warn-on-forecast window while remaining well calibrated. By removing the numerical model from both training and initialization, StormScope sidesteps the data-assimilation and microphysics-approximation bottlenecks that constrain convection-allowing models and their emulators.

The convective-initiation and intensification results provide strong evidence that the model has learned physically meaningful precursors of intense convection. Given intrinsic storm-scale uncertainty, StormScope's generative formulation samples sharp, physically plausible trajectories rather than a blurred mean, providing ensemble spread while preserving the realistic morphology valued by forecasters.

Like ML models for global medium-range weather prediction~\citep{lam2023learning, bonev2023spherical, lang2024aifs, price2025probabilistic, alet2025skillful}, StormScope also delivers an enormous gain in computational efficiency. On 2 NVIDIA B300 GPUs in parallel, StormScope produces a 2-h forecast for the continental United States (Fig.~\ref{fig:intro_combined}a) in about 5.5 minutes, enabling very rapid nowcast updates and large storm-scale ensembles. Significant speedups are possible on smaller domains,  potentially enabling
significant advances in the timeliness of severe weather warnings.

StormScope achieves substantial improvements relative to NWP in thunderstorm nowcasting. For context, recent improvements to ECMWF's ML AIFS model yield global 500-hPa anomaly correlation coefficients (ACC) exceeding those of the NWP IFS model by roughly 4\% and 11\% at forecast lead times of 4 and 8 days \cite{Moldovan_etal_GMD_2026}. For probabilistic forecasting, FGN reduces global 500-hPa geopotential CRPS relative to ECMWF ENS by approximately 20\% and 11\% at these lead times \citep{alet2025skillful}. On metrics which are relevant and widely used for storm-scale nowcasting, StormScope's improvements over NWP are a lot more substantial. For radar echo thresholds of 40~dBZ at 30-min and 120-min lead times, the CSI values for StormScope exceed those for WoFS by 76\% and 45\%, while the corresponding improvements at 55~dBZ are about 200\% and 16\%. The highest-impact, least-well-parameterized scales of weather may be those for which training on observations yields the greatest advance.

StormScope's ensemble forecasts grow progressively underdispersive at longer lead times, and its skill converges toward the operational baseline by 3~h. 
StormScope does not directly predict hazards that warnings target, such as hail-size or wind-speed probabilities. These limitations motivate extending the forecast state to hazard variables such as hail size and wind gusts, conditioning on larger-scale synoptic context to extend skill beyond 2 h, and improving long-lead-time ensemble dispersion. 

\section{Methods}\label{sec:methods}

\subsection{Evaluation and comparison with WoFS}

WoFS generates state-of-the-art fine-scale numerical weather predictions for 18 ensemble members at 5-min temporal and 3-km horizontal resolution on 51 vertical levels. Advanced data-assimilation algorithms ingest data from the GOES satellites and radar reflectivities and velocities. WoFS forecasts are a key tool used by NWS meteorologists to provide short-term guidance for impending severe thunderstorm hazards. To place the two systems on an even footing, we bias-correct WoFS (Extended Data Sec.~\ref{secA:wofs_bias_correction}) and evaluate all qualifying WoFS initializations from 2025 ($n = 929$ forecasts; March--August) on matched domains and verification times. Deterministic skill is assessed using the first indexed ensemble member and neighborhood Critical Success Index (CSI) at 40- and 55-dBZ thresholds with a 33~km neighborhood. An object-based framework is employed to track storm objects and stratify skill by storm age for analyzing skill at forecasting storm intensification and convective initiation. The object-based evaluation independently corroborates the grid-based CSI results (Extended Data Fig.~\ref{fig:combined-2025}). Probabilistic skill is assessed using the probabilistic Fractions Skill Score (pFSS) and ensemble calibration curves with 16-member ensembles for each model. For WoFS, the first 16 of 18 indexed ensemble members were used.

\textbf{Notation.} Throughout the evaluation sections, subscripts $f$ and $v$ denote the forecast and verification sides. $Z_{f,i}$ and $Z_{v,i}$ are the forecast and verification reflectivity fields at grid point $i$; $\mathbbm{1}[\cdot]$ is the indicator function; $B_{f,i} = \mathbbm{1}[Z_{f,i} \geq \tau]$ is the field binarized at the event threshold $\tau$ (40~dBZ unless stated otherwise); and $\mathcal{N}(i)$ is the square neighborhood of width 11 grid cells ($\approx 33$~km) centered on grid point $i$. Ensemble members are indexed by $e$ and written as superscripts, e.g., $B^{(e)}_{f,i}$, with $\langle\cdot\rangle_e$ denoting an average over the ensemble. Contingency counts are hits $H$, misses $M$, false alarms $F$, and extras $E_f$ and $E_v$. In the object-based framework, $A_f$ and $A_v$ are object areas; $t_f$ and $t_v$ are object ages; $\ell$ is the forecast lead time; and $a = t_v - \ell$ is the object age at forecast initialization. The object-detection thresholds are $\tau_{\mathrm{low}}$, $\tau_{\mathrm{high}}$, and $A_{\mathrm{min}}$, and $\delta$ is the temporal matching tolerance.

\subsubsection{Neighborhood contingency evaluation}\label{sec:nbhd-eval}
Exact-pixel verification of reflectivity double-penalizes
small displacement errors: a storm forecast $10$~km from its observed
position counts as both a false alarm and a miss. Neighborhood
contingency relaxes this by asking whether an event occurs \emph{near}
each grid point.

Forecast and verification fields are first binarized at a specified threshold $\tau$ ($B_{f(v),i} = \mathbbm{1}[Z_{f(v),i} \geq \tau]$). The fields are spatially dilated with the same square window $\mathcal{N}(i)$ of width $11$ pixels
($\approx 33$~km) centered on grid point $i$:
\begin{equation}
  \hat{B}_{f,i} \;=\;
    \max_{j \in \mathcal{N}(i)} B_{f,j},
  \qquad
  \hat{B}_{v,i} \;=\;
    \max_{j \in \mathcal{N}(i)} B_{v,j},
\end{equation}

A $2\times 2$ contingency table is accumulated over grid points with valid MRMS observations. A \emph{hit} is recorded where $\hat{B}_{f,i} = \hat{B}_{v,i} = 1$, a \emph{false alarm} where $\hat{B}_{f,i} = 1$ and $\hat{B}_{v,i} = 0$, and a \emph{miss} where $\hat{B}_{f,i} = 0$ and $\hat{B}_{v,i} = 1$. The Critical Success Index follows as

\begin{equation}
  \mathrm{CSI} \;=\; \frac{H}{H + F + M},
\end{equation}

with counts pooled over all cases and forecast frames at each lead time.

\subsubsection{Object-based contingency evaluation}\label{sec:object-based-eval}
We track storm objects and label them according to age for the convective-initiation evaluation of Fig.~\ref{fig:convective_initiation_wofs_ss}. In each frame, forecast and observed cells are identified, matched one-to-one, and tallied as hits, misses, false alarms, and ``extras.'' This follows the Warn-on-Forecast System (WoFS) verification methodology of Ref.~\cite{skinner2018object}, with the same conventions later used by Refs.~\cite{potvin2020assessing, duda2023using, britt2024verification, skinner_2025}. We implement this two-threshold, connected-component, total-interest procedure using \texttt{scikit-image} for connected-component labeling and region properties. Aggregating the same per-frame contingency counts without age stratification yields a conventional object-based skill comparison between StormScope and WoFS, presented in Extended Data Fig.~\ref{fig:combined-2025}.

\paragraph{Two-threshold connected-component identification}
Object identification uses a two-threshold rule so that a cell must contain a strong core but may extend to a weaker boundary. The frame is binarized using a threshold mask to retain all pixels at or above $\tau_{\mathrm{low}} = 40$~dBZ. The binarized frame is then grouped into connected components using queen adjacency implemented through \texttt{skimage.measure.label}. Each component is a candidate cell. A candidate cell is retained only if it satisfies two conditions: its area must be greater than 108~km$^2$, and the maximum reflectivity within the candidate cell must be at least $\tau_{\mathrm{high}} = 45$~dBZ. The surviving cells are labeled as storm objects, and the rejected cells are zeroed out.

\paragraph{Total-interest matching}

Given the forecast cell set and the verification cell set for a frame, every
candidate (forecast, verification) pair is scored by a total interest (TI) score
combining centroid agreement and boundary agreement. For each forecast–verification object pair, let $d_c$ denote the distance between the object centroids and $d_b$ the minimum distance between their boundaries. The corresponding distance cutoffs, $D_c$ and $D_b$, are both set to \SI{40}{\km}:
\begin{align}
  s_c &= \max\!\left(0,\ \frac{D_c - d_c}{D_c}\right), &
  s_b &= \max\!\left(0,\ \frac{D_b - d_b}{D_b}\right), &
  \mathrm{TI} &= \tfrac{1}{2}\,(s_c + s_b).
\end{align}

An area-ratio weight ($w$) penalizes size mismatch:

\begin{align}
  w = \frac{\min(A_f, A_v)}{\max(A_f, A_v)}, \qquad
  \mathrm{TI}_w = w \cdot \mathrm{TI},
\end{align}
where $A_f$ and $A_v$ are the forecast and verification cell areas.

An iterative algorithm matches verification objects to forecast objects using $K = 5$ passes. First, the TI score is computed for all ordered pairs of forecast and verification objects. At the beginning of each iteration, all objects are set to unmatched. For every iterative pass, each unmatched verification cell is provisionally paired with its best unmatched forecast partner according to the highest weighted total-interest score $\mathrm{TI}_w$; when two verification cells claim the same forecast cell, the pair with the higher $\mathrm{TI}_w$ wins, and the loser returns to the unmatched pool for the next pass. A provisional pair is accepted only if its \emph{plain} total interest strictly exceeds the canonical threshold $\mathrm{T} = 0.2$ (i.e., $\mathrm{TI} > \mathrm{T}$).

\paragraph{Contingency classification}

Matched pairs are hits. Unmatched forecast objects with a maximum possible $\mathrm{TI}$ among all verification objects less than or equal to $\mathrm{T}$ are false alarms. Unmatched verification objects with $\max(\mathrm{TI}) \leq \mathrm{T}$ are misses. All other unmatched forecast and verification objects (those with a maximum plain $\mathrm{TI}$ above $\mathrm{T}$ that nonetheless lost the one-to-one assignment) are considered ``extras.'' The distinction between an extra and a false alarm is whether a real storm underlies the object: a false alarm has no observed counterpart within the matching tolerance and represents genuine overprediction, whereas an extra overlaps a real storm and is a redundant (e.g., fragmented or split) object. Extras are therefore excluded from POD, SR, and CSI so that these skill scores reflect event detection.

\begin{align}
  \mathrm{POD} &= \frac{H}{H + M}, &
  \mathrm{SR} &= \frac{H}{H + F}, &
  \mathrm{CSI} &= \frac{H}{H + F + M}, &
  \mathrm{FB} &= \frac{H + F + E_f}{H + M + E_v},
  \label{eqn:contingency}
\end{align}

where $E_f$ and $E_v$ are forecast and verification ``extras'', and FB denotes frequency bias. The counts $H$, $F$, and $M$ and the CSI are parallel to those of the neighborhood evaluation, with objects in place of dilated pixels.

\paragraph{Object age tracking and the age-at-initialization coordinate}

To stratify skill by convective maturity, we track objects through consecutive verification (and forecast) frames using grid-cell overlap: an object inherits the age of the oldest preceding-frame object whose footprint it overlaps (by at least one grid cell), incremented by the elapsed frame interval, and is assigned age zero at first detection when it overlaps no prior object. Verification-side ages ($t_v$, from a continuous observed-radar record with an extended pre-forecast warm-up window) and forecast-side ages ($t_f$, grown from forecast initialization) are tracked independently.

We report metrics on the \emph{(object age at initialization $\times$ forecast lead)} plane. For a pair verifying at lead $\ell$ (minutes), the age at initialization is
\begin{align}
  a = t_v - \ell \quad\text{(truth-side objects)}, \qquad a = t_f - \ell \quad\text{(forecast-side, false alarms only)}.
\end{align}
Objects with $a \geq 0$ existed at forecast initialization (the \emph{pre-existing} regime); objects with $a < 0$ formed after initialization (the \emph{convective-initiation}, or CI, regime). Hits and misses are binned by their observed age-at-initialization; in the CI regime, false alarms are binned by their forecast age-at-initialization. Extras are excluded.

\paragraph{Temporal-neighborhood matching in the convective-initiation regime}

Having an age for every object also lets us sharpen the definition of a hit for storms that are born inside the forecast window. Observations, never forecasts, define the space--time location $(x, y, t)$ of a true event. Accordingly, in the CI regime, we require a forecast object to match an observed storm in both space \emph{and} time: a candidate pair is accepted as a hit only if it is spatially colocated (plain $\mathrm{TI} > \mathrm{T} = 0.2$, as above) \emph{and} temporally consistent, i.e., their ages agree within a tolerance $\pm\delta$ ($\left| t_f - t_v \right| \leq \delta$, with a default $\delta = 10$~min matching the 10-min forecast cadence). This test is well posed in the CI regime because both the observed and forecast storms are born inside the forecast window, so their ages are uncensored and directly comparable. The total-interest score of any spatially plausible but temporally inconsistent pair is zeroed \emph{before} the iterative assignment, so such a pair does not become a hit but instead falls through to a false alarm (forecast side) and a miss (verification side).

Pre-existing storms ($a \geq 0$) are matched on space alone: a persisted forecast object has no pre-initialization history, so its age cannot be compared against the age of an older observed storm. We therefore do not score false alarms by age in the pre-existing regime; only POD---which is defined purely from observed hits and misses---is reported there.

\subsubsection{Probabilistic Fractions Skill Score}\label{sec:pfss}
The probabilistic Fractions Skill Score (pFSS)~\citep{necker} generalizes the Fractions Skill Score for an ensemble forecast. First, the forecast or observation field is binarized at a given threshold $\tau$ as

\begin{align}
    B_{f(v),i} &= \mathbbm{1}\!\left[Z_{f(v),i} \geq \tau\right].
\end{align}

The neighborhood probability averages the binarized field over the neighborhood $\mathcal{N}(i)$ centered on grid point $i$:
\begin{align}
    \mathrm{NP}_{f,i} = \langle B_{f,j} \rangle_{j \in \mathcal{N}(i)},
\end{align}
and likewise $\mathrm{NP}_{v,i}$ for the verification field. The neighborhood ensemble probability is the mean of the neighborhood probabilities across ensemble members. Denoting member probability as $\mathrm{NP}^{(e)}_{f,i}$,
\begin{align}
    \mathrm{NEP}_{f,i} = \langle \mathrm{NP}^{(e)}_{f,i} \rangle_{e \in \mathrm{ensemble}}.
\end{align}
Finally, pFSS is defined as
\begin{align}
    \mathrm{pFSS} = 1 - \frac{\sum_i \left( \mathrm{NEP}_{f,i} - \mathrm{NP}_{v,i} \right)^2}{\sum_i\mathrm{NEP}^2_{f,i} + \sum_i \mathrm{NP}_{v,i}^2}.
\end{align}
Here, the summation $\sum_i$ runs over all valid grid points $i$.

\subsubsection{Ensemble Reliability Diagrams}\label{sec:calibration}
Neighborhood event fractions average the binarized fields over the same $11 \times 11$-grid-point square unless otherwise specified.
\begin{align}
    \tilde{B}_{\cdot, i} &= \langle B_{\cdot, j} \rangle_{j \in \mathcal{N}(i)}.
\end{align}
The forecast event probability and the observed neighborhood fraction are $p_i = \langle\tilde{B}^{(e)}_{f,i}\rangle_e$ and $Y_i = \tilde{B}_{v,i}$.

Forecast--observation pairs $(p_i, Y_i)$ are accumulated in $K=101$ probability bins centered on $c_k = k/(K-1)$, $k = 0, \dots, K-1$. Each valid grid point is assigned to the bin nearest $p_i$. Forecast counts, forecast-probability sums, and observed-fraction sums are accumulated over the domain and across cases, stratified by lead time. 

Before plotting, fine bins are merged into a small number of approximately equally populated bins~\citep{broecker2007}. Each merged bin $B_k$ has an adaptive population target and a minimum forecast count. We also cap the probability span of any given bin at 0.1 since probabilities are strongly skewed towards zero and a pure equal-population rule would collapse the sparse upper range into a misleadingly broad bin. This construction balances statistical support and resolves the probability distribution.

Each merged bin contributes one reliability point $(r_k, f_k)$, where
\begin{align}
    r_k &= \frac{1}{n_k} \sum_{i \in B_k} p_i,
    &
    f_k &= \frac{1}{n_k} \sum_{i \in B_k} Y_i,
\end{align}
with $n_k$ the number of forecasts in the bin. The point is placed at the mean forecast probability $r_k$ of the samples actually contained in the bin rather than at the bin midpoint; with unequal bin widths the two can differ substantially, and for a reliable forecasting system the expected observed frequency corresponds to $r_k$ \citep{broecker2007}.

Curves along $f_k = r_k$ indicate well-calibrated forecasts. A curve below the diagonal indicates overconfidence, consistent with an underdispersive ensemble; a curve above the diagonal indicates an overdispersive, underconfident model.

\subsubsection{Evaluation set}

StormScope was trained on MRMS data from 2020--2023 and GOES data from 2018--2023. The year 2024 was used for validation and checkpoint selection. All evaluation was performed in the held-out year 2025. Of 951 WoFS forecasts across different domains available from the National Severe Storms Laboratory (NSSL), 22 had an irregular number of forecast frames (rather than the canonical WoFS cadence of 5-min time steps) or were missing key metadata fields and were excluded. All remaining 929 forecasts were used. StormScope forecasts were generated at matching initialization times. StormScope forecasts the entire CONUS, while WoFS generates forecasts on roughly 900~km~$\times$~900~km domains. StormScope forecasts were extracted from the full-CONUS forecast on domains matching the WoFS domain, and evaluation was performed at matching lead times.

\subsubsection{Numerical benchmarks}\label{sec:numerical-benchmarks}
Extended Data Table~\ref{tab:wofs_vs_stormscope_skill} reports POD and SR for the neighborhood-grid evaluation.

Extended Data Table~\ref{tab:wofs_vs_stormscope_skill} shows numerical values of object-based CSI, neighborhood-grid CSI, and 16-member ensemble pFSS for WoFS and StormScope. These numbers correspond to Fig.~\ref{fig:combined_skill_figure}a,b and Extended Data Fig.~\ref{fig:combined-2025}a.

\subsection{Hazardous Weather Testbed Spring Forecasting Experiment}
The Hazardous Weather Testbed (HWT), at the National Weather Center in Norman, Oklahoma, provides a collaborative environment for conducting forecasting experiments that evaluate emerging concepts, tools, and algorithms developed by NOAA’s National Severe Storms Laboratory (NSSL), Storm Prediction Center (SPC), and collaborating institutions. Established in 2000, the Spring Forecasting Experiment (SFE) is the HWT’s longest-running experiment and is held annually during the height of the severe weather season.

The SFE is jointly led by SPC and NSSL and is designed to facilitate the transition of research into operational forecasting. This work includes evaluating innovative severe weather prediction tools and forecasting techniques, examining how end users interpret and apply severe weather guidance, and investigating convection-allowing model (CAM) ensemble configurations that can contribute to the development of NOAA’s Unified Forecast System (UFS; https://ufs.epic.noaa.gov/). The concentration of severe weather forecasting and research expertise at the National Weather Center, together with advanced visualization capabilities, carefully designed experimental frameworks, and extensive collaborations, has established the annual SFE as a highly influential and respected severe weather forecasting research effort.

\subsection{Data}
\label{sec:data}

The model is trained solely on observations from the GOES-16 satellite and the Multi-Radar/Multi-Sensor (MRMS) system. The model was trained on GOES data from 2018 to 2023 and MRMS data from 2020 to 2023.

\begin{itemize}
    \item \textbf{Satellite data:} We use 8 GOES-16 ABI channels (Table~\ref{tab:data}) regridded to a \qty{3}{\kilo\metre} Lambert Conformal Conic grid at a 10-minute temporal cadence. Lightning observations are taken from the Geostationary Lightning Mapper Level-2 event product. We extract latitude and longitude of detected events. Events are accumulated into 10-minute windows onto a regular 0.1° latitude–longitude grid. The counts are transformed as $\log(1 + x)$ to compress their heavy-tailed distribution and bilinearly interpolated onto the shared 3~km grid. No satellite parallax corrections are applied.
    \item \textbf{Radar data:} MRMS products provide the ground-truth base and composite reflectivity interpolated to the same \qty{3}{\kilo\metre} grid at a 10-minute cadence.
\end{itemize}

\begin{table}[t]
\centering
\resizebox{\linewidth}{!}{%
\begin{tabular}{llll}
\toprule
Variable                 & Description                            & Modality       & Use \\
\midrule
Blue                     & Visible band at \qty{0.47}{\micro\metre}                & Satellite & I, F \\
Red                      & Visible band at \qty{0.64}{\micro\metre}                & Satellite & I, F \\
Veggie                   & Near-IR band at \qty{0.86}{\micro\metre}                & Satellite & I, F \\
Shortwave window         & IR band at \qty{3.90}{\micro\metre}                     & Satellite & I, F \\
Upper-level water vapor  & IR band at \qty{6.19}{\micro\metre}                     & Satellite & I, F \\
Mid-level water vapor    & IR band at \qty{6.95}{\micro\metre}                     & Satellite & I, F \\
Lower-level water vapor  & IR band at \qty{7.34}{\micro\metre}                     & Satellite & I, F \\
Clean IR longwave window & IR band at \qty{10.35}{\micro\metre}                    & Satellite & I, F \\
Aggregated lightning density & Geostationary Lightning Mapper event counts                  & Satellite & I, F \\
Radar                    & MRMS maximum composite reflectivity & MRMS radar     & I, F \\
Radar                    & MRMS base reflectivity & MRMS radar     & I, F \\
Time/solar illumination & Cosine of solar zenith angle & Computed & I \\
Topography & Elevation above sea level & Static & I \\
Radar coverage & Sum of Gaussians at radar locations & Computed/static & I \\
\bottomrule
\end{tabular}
}
\caption{Summary of input (I) and forecast (F) data streams.}
\label{tab:data}
\end{table}

\subsection{Model details}

The system consists of two coupled Diffusion Transformer (DiT) branches, each a purely data-driven generative model conditioned on a temporal window of recent observations. A satellite branch forecasts the eight GOES ABI channels. A radar branch forecasts MRMS composite and lowest-altitude reflectivity together with GLM lightning-event counts accumulated over 10-min windows and is jointly conditioned on the satellite, radar, and lightning histories. Forecasts are produced with an autoregressive sliding window: at each \qty{10}{\minute} step, the branches predict the single next state, which is appended to the conditioning window while the oldest frame is discarded, so that predicted states are iteratively recycled as inputs for subsequent steps. This enables the model to generate continuous forecasts across extended lead times while maintaining physical consistency between the multimodal fields. The satellite branch is functionally independent of the radar branch and can be trained and sampled on its own; the radar branch consumes the satellite branch's predictions as its satellite conditioning window rolls beyond the available observations.

\subsubsection{Neural network architecture}

The model uses a Diffusion Transformer (DiT) architecture~\cite{peebles2023scalable} adapted to large two-dimensional geophysical spatial fields using efficient local attention~\cite{hassani2023neighborhood,hassani2024faster,hassani2025generalized}. Each network operates on a $1024 \times 1792$ Lambert Conformal Conic grid. The six most recent states of each observed modality and the exogenous variables---normalized latitude and longitude, the cosine of the solar zenith angle at the input and target times, a NEXRAD radar-proximity (coverage) map, and topography---are concatenated channel-wise with the noisy latent state (75 input channels for the radar branch and 61 for the satellite branch). The model inputs are first fed through a standard strided convolution-based patch embedding, with a patch size of $4 \times 4$, to produce a sequence of $256 \times 448 = 114{,}688$ tokens with embedding dimension 768, to which a learnable positional embedding is added. These are then processed by 16 transformer blocks with 12 attention heads, which are conditioned on the diffusion noise level using the standard adaptive layer normalization (adaLN-Zero) adopted in DiTs. To mitigate the computational complexity of standard all-to-all attention at this token count, we use sparser two-dimensional Neighborhood Attention (NA)~\citep{hassani2023neighborhood, hassani2024faster, hassani2025generalized}. Unlike global attention, NA restricts each token's receptive field to a local window of size $49 \times 49$, greatly reducing the computation required for self-attention while allowing long-range dependencies to be captured gradually over multiple layers. The final prediction is produced by taking the output of the DiT blocks, applying an adaLN-modulated linear projection, and reshaping back to the $1024 \times 1792$ grid.

Each network has 261 million parameters. Each expert is trained on 32 NVIDIA H100 GPUs arranged as 16 data-parallel groups with two-way spatial-domain parallelism within each group: activations are sharded along the grid-height dimension, and the neighborhood-attention halos are exchanged between shards (Extended Data Sec.~\ref{secA:domain_parallel}). Domain parallelism keeps the per-GPU memory footprint tractable at \qty{3}{\kilo\metre} resolution and provides a direct path to training on still larger grids.

\subsubsection{Training and inference}
We formulate nowcasting as learning the conditional distribution of the future atmospheric state $x_{t+}$ given a sequence of recent observations $x_{t-}$. We adopt the Elucidated Diffusion Model (EDM) framework~\citep{karras2022elucidating}, which provides a well-behaved discretization of the diffusion stochastic differential equation (SDE). The network learns $p(x_{t+} \mid x_{t-}; c)$, where $x_{t+}$ is the state one \qty{10}{\minute} step ahead, $x_{t-}$ is a temporal stack of the six most recent states (spaced 10~min apart over the preceding hour), and $c$ denotes static and dynamic
conditioning (e.g., topography and the solar zenith angle derived from the time of day). The denoiser $\mathcal{D}_\theta$ is trained to minimize the EDM-weighted denoising loss:
\begin{equation}
    L(\theta) = \mathbb{E}_{x, \sigma, n} \left[ \lambda(\sigma)
    \lVert \mathcal{D}_\theta(x_{t+} + n_\sigma; x_{t-}; c; \sigma) - x_{t+}
    \rVert^2_2 \right],
\end{equation}
where $n_\sigma \sim \mathcal{N}(0, \sigma^2 I)$ and $\lambda(\sigma)$ is the standard EDM weighting. EDM input and output preconditioning keeps signals at bounded variance across the full range of noise levels $\sigma$. Grid points with missing observations (e.g., outside radar coverage) are masked out of the loss.

To capture both large-scale organization and fine-scale convective detail, we adopt a multi-expert denoising strategy~\citep{balaji2022ediff, brenowitz2025climate}. Denoising networks conventionally share a single set of parameters across the entire range of noise levels. However, both the character of the denoising task and the amount of useful training signal vary strongly with $\sigma$. At high noise levels, the network need only recover the large-scale structure of the target, which is tightly constrained by the prior-step conditioning and changes little over a single 10-min model step; this regime is comparatively easy and can be learned from relatively few samples. We observe characteristics of overfitting if the high-noise expert is trained longer. At low noise levels, by contrast, the network must reconstruct the fine storm structure and morphology that govern the evolution of storm-scale dynamics. Assigning a dedicated expert to each regime allows each network to specialize rather than compromise across these competing objectives. We partition the noise range $[\sigma_{\min}, \sigma_{\max}]$ into three regimes and train a separate expert on each, drawing the training noise uniformly in $\log\sigma$ within an expert's assigned range; the experts span $\sigma \in [0.005, 3]$, $[2, 100]$, and $[50, 800]$. The low-, mid-, and high-noise experts are trained for progressively fewer epochs (20, 10, and 5, respectively).

Each expert is trained independently with a global batch size of 16.
The learning rate follows a cosine-annealing schedule with warm restarts from a peak of $5 \times 10^{-4}$: each cycle spans one-quarter of the scheduled iterations, the peak learning rate is halved after every restart, and the rate decays to a floor of $1 \times 10^{-5}$ within each cycle.

At inference, we use the second-order (Heun) stochastic sampler of Ref.~\cite{karras2022elucidating} with noise churn ($S_{\mathrm{churn}} = 10$),
integrating from $\sigma_{\max} = 800$ to $\sigma_{\min}$ ($0.005$ for
radar, $0.001$ for the satellite) in 100 steps at maximum
fidelity using $\rho = 7$ discretization. A deterministic router selects experts at predefined boundaries $\sigma \in \lbrace 3, 100 \rbrace$ as the noise schedule descends. Within a forecast step the satellite branch is sampled first and the radar branch is
then sampled conditioned on the updated satellite window, so multimodal
forecasts run the two branches in sequence. Sampling is executed under
\texttt{bfloat16} autocast with CUDA graph compilation
(\texttt{torch.compile}), and ensemble forecasts are generated by drawing
independent initial latent noise per member (16 members in our evaluation
configuration). Inference could be further accelerated by
techniques such as distillation, domain parallelism, and/or reduced-precision inference.

\backmatter

\bmhead{Acknowledgments}
We thank our NVIDIA colleagues Julius Berner, Arash Vahdat, Jean Kossaifi, and Linnea Wolniewicz for helpful discussions. We also thank Tom Hamill and Montgomery Flora at The Weather Company for their helpful comments on early versions of this work. We thank John Cintineo at NSSL for feedback. Funding for NSSL/NOAA coauthors was provided by NOAA/OAR under the NOAA-University of Oklahoma Cooperative Agreement NA21OAR4320204. Work by federal employees was conducted as part of their regular duties. The scientific results, conclusions, and views expressed herein are those of the authors and do not necessarily reflect those of OAR or the Department of Commerce.

\bmhead{Data Availability}
The trained model weights are available at \url{https://huggingface.co/nvidia/stormscope-goes-mrms/}. Example inference code is available at \url{https://github.com/NVIDIA/earth2studio/tree/b433a69608cbe9db22085ec72d17da93164e5dfa/examples/04_nowcasting}.

\bmhead{Use of Large Language Models}
Large language models were used to assist with software development, copyediting and refinement of the manuscript. All LLM outputs were verified carefully by the authors. The authors take full responsibility for the accuracy and integrity of the manuscript, code, results and conclusions.

\clearpage

\begin{appendices}

\section{Object-based skill comparison between StormScope and WoFS}\label{secA:objcomp}
\begin{figure}[H]
    \centering
    \includegraphics[width=0.8\linewidth]{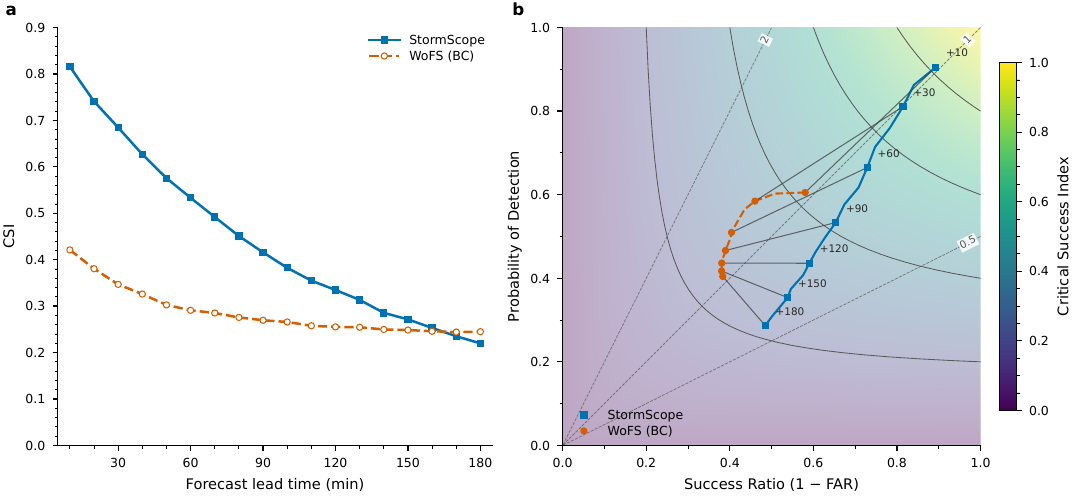}
    \caption{\textbf{Object-based deterministic comparison of bias-corrected
    Warn-on-Forecast System (WoFS) and StormScope.}
    All comparisons use identical spatial domains and matched
    verification timestamps; WoFS composite reflectivity is bias-corrected per lead time. The evaluation set comprises all WoFS initializations run from March through August 2025 (929 of 951 forecasts were used after excluding runs with a noncanonical forecast cadence or missing data).
    \textbf{a}, Object-based Critical Success Index (CSI) as a
    function of forecast lead time. Storm cells are identified and matched using the methodology described in Sec.~\ref{sec:methods} with dual 40-dBZ/45-dBZ minimum/core thresholds. Both forecasts are scored at ensemble member~0.
    \textbf{b}, Roebber performance diagram derived from the
    object-based contingency table of panel~\textbf{a}. Each marker is one
    forecast lead, plotted at the pooled Success Ratio and Probability of Detection (POD).
    For readability, markers and lead
    labels are placed only at $+10$, $+30$, $+60$, $+90$, $+120$,
    $+150$, and $+180\,\mathrm{min}$, and at each of these leads a
    thin gray tie-line joins the corresponding StormScope and
    WoFS~(BC) markers. Background shading is the iso-CSI field.
    Thin dashed gray diagonals mark the frequency-bias isolines.}
    \label{fig:combined-2025}
\end{figure}

The object-based storm-tracking framework (Methods Sec.~\ref{sec:object-based-eval}), whose primary role in this work is the age-stratified convective-initiation analysis, also provides an independent, storm-centric check on the grid-based neighborhood skill reported in the main text. Extended Data Fig.~\ref{fig:combined-2025} compares bias-corrected WoFS, hereafter WoFS (BC), and StormScope using the object-based CSI and the associated Roebber performance diagram; numerical values are tabulated in Extended Data Table~\ref{tab:wofs_vs_stormscope_skill}. The object-based comparison shows the same magnitude and lead-time dependence of the StormScope advantage as the grid-based evaluation.

\section{Statistical comparison of paired forecasts}\label{secA:statcomp}

\begin{figure}[H]
    \centering
    \includegraphics[width=\linewidth]{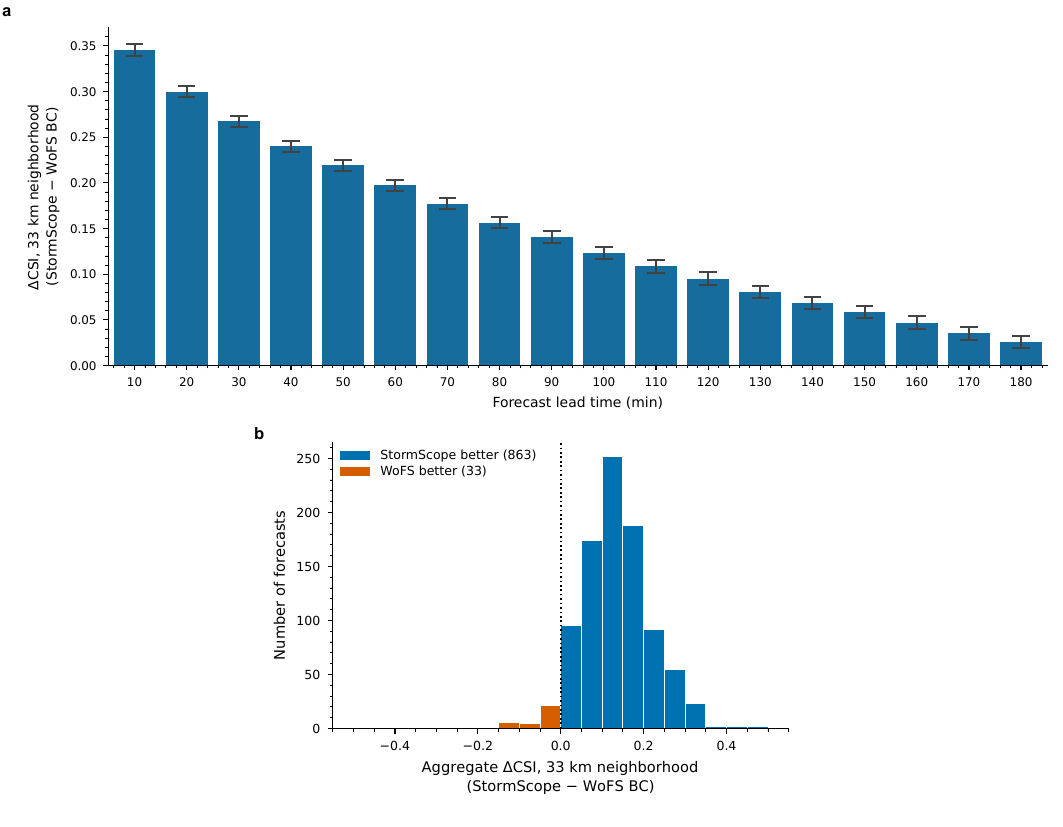}
    \caption{\textbf{Paired forecast skill comparison.} \textbf{a}, Mean within-initialization difference in neighborhood-grid CSI (StormScope--WoFS (BC)) as a function of forecast lead time, with 95\% bootstrap confidence intervals ($n = 929$ paired 2025 forecasts). \textbf{b}, Distribution of the per-forecast difference in trajectory-aggregated CSI (StormScope--WoFS (BC)); among the 918 forecasts with valid 40-dBZ cells, StormScope attains the higher aggregate CSI in 863 cases (94\%), WoFS does so in 33 cases (3.6\%), and 22 cases (2.4\%) are tied. In \textbf{a}, blue bars show mean differences and gray whiskers show 95\% confidence intervals. In \textbf{b}, blue and orange bars indicate higher skill for StormScope and WoFS; the black dotted vertical line marks equal skill.}
    \label{fig:paired-csi}
\end{figure}

We present further details on the statistical significance of the comparison between WoFS and StormScope. Figure~\ref{fig:paired-csi}a shows the differences in CSI across lead times for paired forecasts made by StormScope and WoFS. For each forecast lead, the bar height is the mean within-initialization difference in skill. The whiskers show a 95\%
nonparametric bootstrap confidence interval for that mean, computed using the \texttt{seaborn} package's \texttt{barplot} tool with \texttt{errorbar=(ci,~95)}, \texttt{n\_boot=10{,}000}, and \texttt{seed=0}.

For a rough aggregate CSI ranking, we collapse the per-lead neighborhood-grid verification of Methods Sec.~\ref{sec:nbhd-eval} to a single skill value per forecast: for each case and each model, we pool the hits, false alarms, and misses across all 18 lead times ($+10$ to $+180$~min at 10-min spacing) and compute one CSI score at 40~dBZ using a 33~km neighborhood and the common WoFS verification-domain mask. Of the 929 scored 2025 WoFS-initialized forecasts, 918 yield a well-defined CSI for both models; the remaining 11 are excluded because at least one model produced no qualifying $\geq 40$~dBZ pixels anywhere in the forecast. Across these 918 forecasts, StormScope attains the higher aggregated CSI in 863 cases (94\%), WoFS does so in 33 cases (3.6\%), and 22 cases (2.4\%) are tied (Fig.~\ref{fig:paired-csi}b).

\section{Additional case studies of severe convective events}\label{secA:cases}

\begin{figure}[H]
    \centering
    \includegraphics[width=0.8\linewidth,height=0.68\textheight,keepaspectratio]{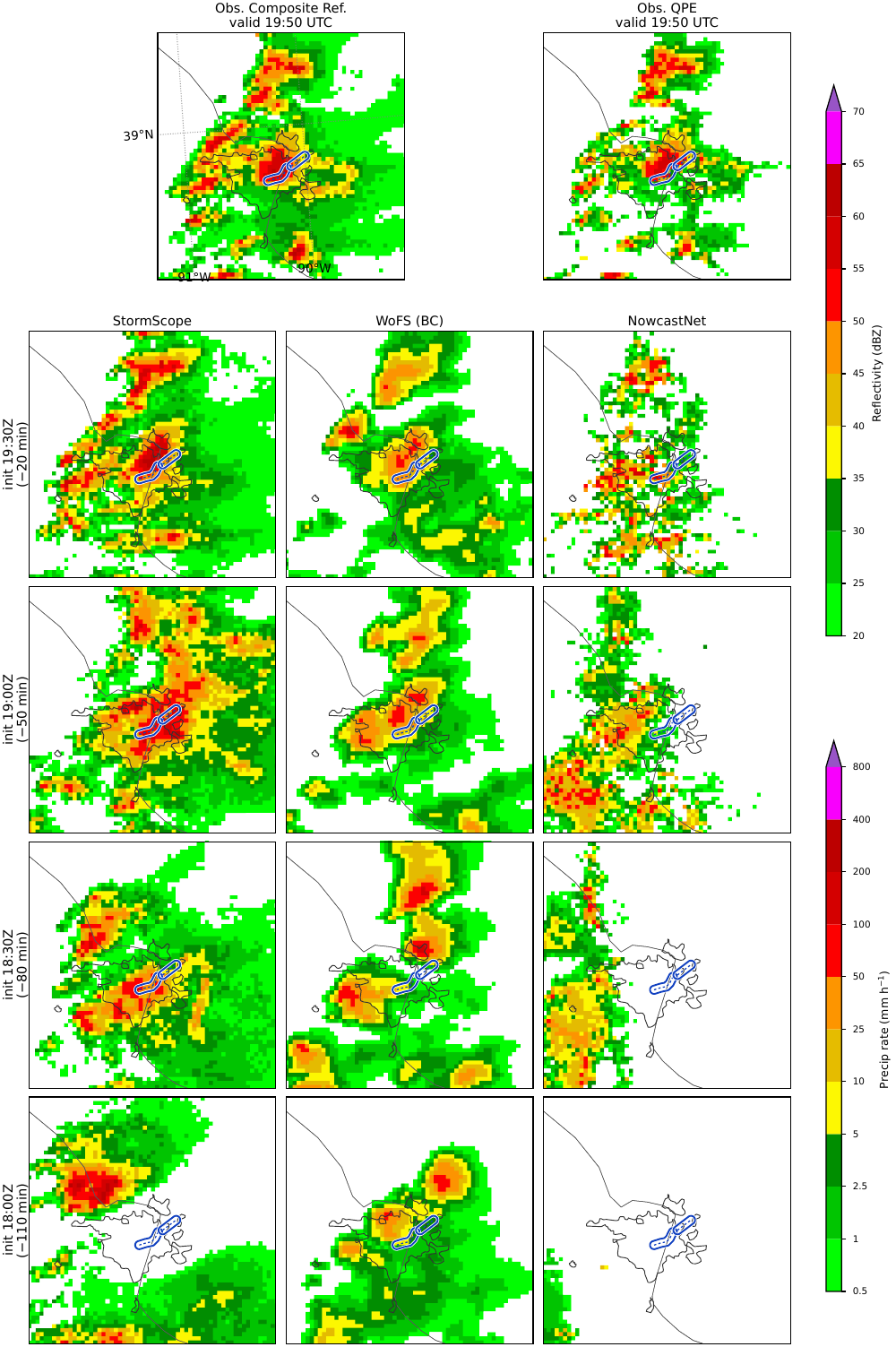}
    \caption{Radar reflectivity verification for the 16 May 2025 St. Louis supercell. Blue outline shows the track of an EF-3 tornado from Clayton through north St. Louis to Granite City, IL between 19:39 and 19:53 UTC~\cite{nws_lsx_2025_stl_tornado}. NowcastNet completely fails to capture the heavy precipitation from the parent supercell.  The St. Louis urban area is annotated for reference.}
    \label{fig:stl}
\end{figure}

\begin{figure}[H]
    \centering
    \includegraphics[width=0.8\linewidth,height=0.72\textheight,keepaspectratio]{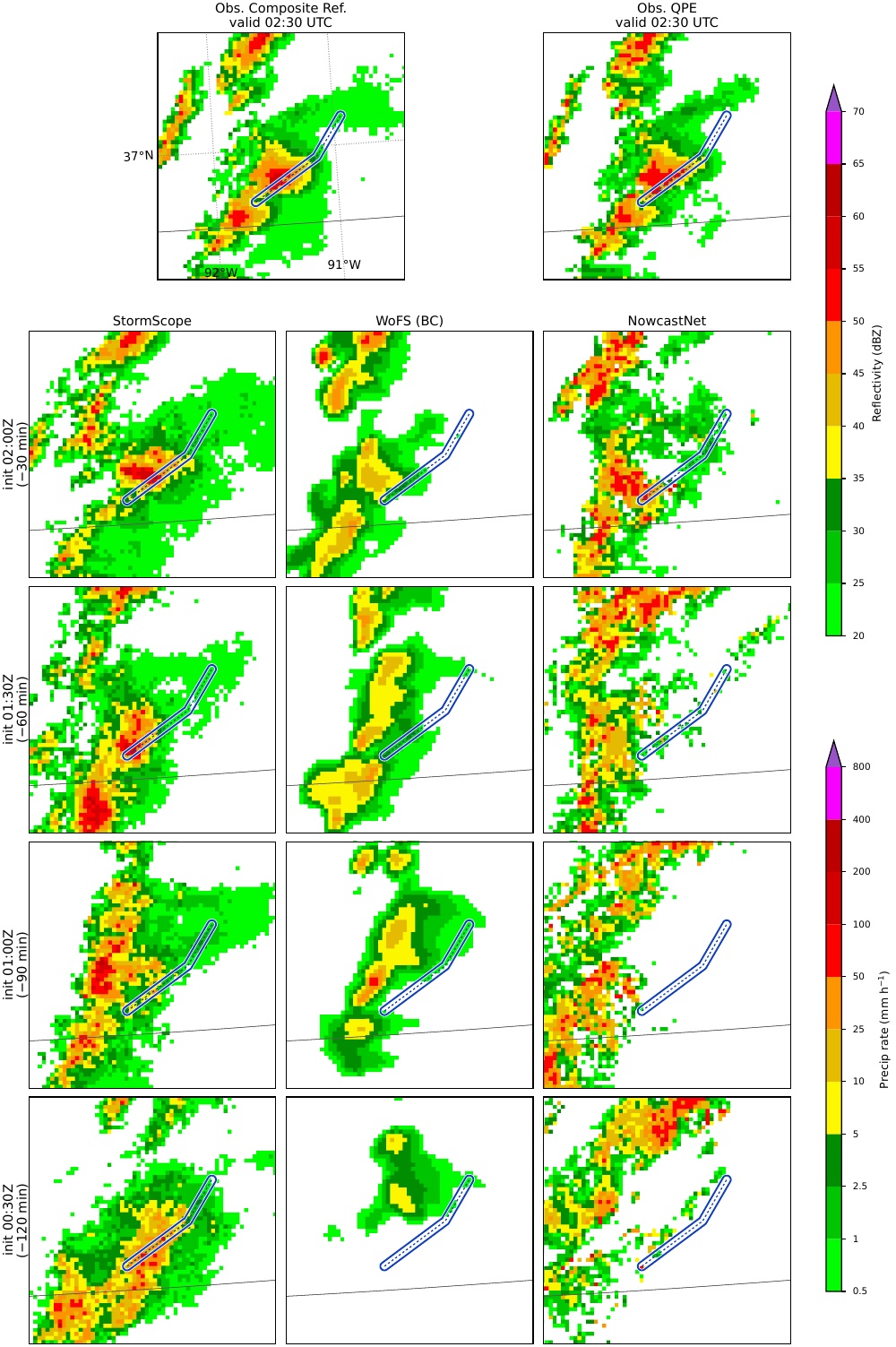}
    \caption{Verification at 02:30 UTC on 15 March 2025.~\cite{nws_sgf_2025_march14} }
    \label{fig:line}
\end{figure}

Examination of additional high-impact case studies confirms that StormScope consistently produces more accurate track and intensity forecasts of extreme thunderstorm events than WoFS or NowcastNet. Two tornadic supercell cases are presented in Figs.~\ref{fig:stl}--\ref{fig:line}: an EF-3 tornado that struck St. Louis, Missouri, on 16 May 2025; and a long-track EF-3 tornado in southeastern Missouri on 15 March 2025.

StormScope shows clear advantages over WoFS and NowcastNet in reproducing observed storm structure across both events and at every lead time considered. WoFS's lower effective resolution smooths the fine-scale storm structures important for supercell hazard prediction and tends to merge multiple ongoing storms into a single cell. NowcastNet's rainfall-rate forecasts are excessively noisy across both cases and all lead times and frequently lack coherent storm structure, limiting their utility for predicting convective storm hazards. Although some small-scale noise is present in StormScope forecasts, they maintain substantially higher fidelity to observed storm structure and intensity in both cases.

StormScope also achieves improved spatial accuracy in forecasting the tornado-producing storms in each case. For the St. Louis supercell, StormScope and WoFS produce qualitatively similar forecasts, while NowcastNet develops a progressively larger westward location bias with increasing lead time, reaching location errors of roughly 100 km at 110 minutes before the event. For the 15 March supercell, StormScope provides forecast improvements, though its cellular storm structure becomes less coherent at 90- and 120-minute lead times. In this case, WoFS forecasts issued 90 and 120 min before verification precede intensification of the tornado-producing cell, and limitations in numerical weather prediction initialization are apparent as a failure to maintain an intense cell in the forecast, with a persistent low bias in predicted reflectivity present in shorter lead-time forecasts. NowcastNet forecasts demonstrate an increasing westward bias in storm location and trend towards incoherency.

\newpage

\section{Extrapolation Baselines}\label{secA:extrapolation}

\begin{figure}[H]
    \centering
    \includegraphics[width=\linewidth]{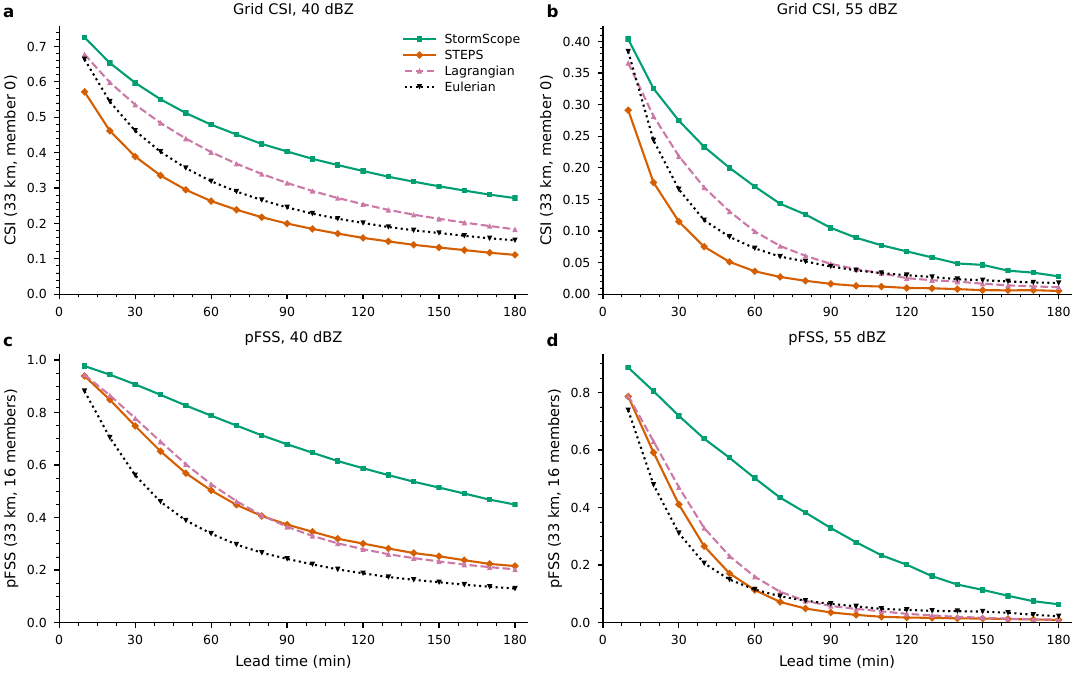}
    \caption{\textbf{StormScope compared with various extrapolation techniques.} Panels (a)-(d) compare StormScope (green solid, squares) with Eulerian (persistence; black dotted, downward triangles), Lagrangian (pink dashed, upward triangles), and STEPS (vermilion solid, diamonds) extrapolation schemes across deterministic 33-km neighborhood CSI (Panels (a) and (b)) and probabilistic FSS (Panels (c) and (d)) at intensity thresholds of 40 dBZ and 55 dBZ.}
    \label{fig:interpolation}
\end{figure}

For completeness, we include a comparison of StormScope forecasts against numerical extrapolation baselines. We benchmark StormScope forecasts against extrapolation methods following the same evaluation protocol as used for other quantitative comparisons (described in Methods) using the same set of 929 forecasts. We use three extrapolation methods: Eulerian persistence, Lagrangian persistence, and the STEPS stochastic nowcasting ensemble, the latter two computed with the open-source pysteps library (v1.21.5)~\citep{pulkkinen2019pysteps}. All baselines operate on the same MRMS composite reflectivity observations that initialize StormScope. The four most recent observations ending at the initialization time ($t-30$, $t-20$, $t-10$, $t_0$ minutes) serve as input to the pysteps protocols. To prevent advection from importing boundary artifacts, the baselines are computed on the forecast domain padded by 96 grid points ($\approx$\qty{290}{\kilo\metre}). All models are verified on identical grids against identical MRMS truth. Each baseline produces 18 steps at a 10-min cadence, matching the lead times of StormScope forecasts. The quantitative skill comparisons between StormScope and extrapolation techniques across neighborhood CSI and pFSS at 40 dBZ and 55 dBZ are shown in Fig.~\ref{fig:interpolation}. We note that StormScope outperforms extrapolation techniques at all lead times and intensity thresholds.

\paragraph{Eulerian Persistence}
The reflectivity field at initialization time is held fixed and serves as a zero-skill forecast at all lead times. 

\paragraph{Lagrangian Persistence}
A dense advection field is estimated from the four input frames with the Lucas-Kanade optical-flow method in the default configuration from pysteps (\texttt{pysteps.motion.get\_method("LK")}). The final observed field is advected forward with the semi-Lagrangian backward scheme of Germann and Zawadzki~\cite{germann2002scale} (\texttt{pysteps.nowcasts.get\_method("extrapolation")}, \texttt{extrap\_method="semilagrangian"}).

\paragraph{STEPS ensemble}
A 16-member stochastic ensemble is generated with
the pysteps implementation of the Short-Term Ensemble Prediction System~\cite{bowler2006steps,pulkkinen2019pysteps} (\texttt{pysteps.nowcasts.get\_method("steps")}), using the same Lucas--Kanade motion field. The reflectivity field is decomposed into eight
spatial cascade levels (\texttt{n\_cascade\_levels=8}), each evolved with a second-order autoregressive model (default \texttt{ar\_order=2}); as the small scales lose predictability they are progressively replaced by spatially correlated stochastic noise (default \texttt{noise\_method="nonparametric"}), and ensemble spread is augmented by stochastic velocity perturbations (default \texttt{vel\_pert\_method="bps"}~\citep{bowler2006steps}). STEPS is run directly in dBZ with a no-echo threshold of -10 dBZ (\texttt{precip\_thr=-10.0}), grid and cadence parameters \texttt{kmperpixel=3.0} and \texttt{timestep=10}, semi-Lagrangian extrapolation (\texttt{extrap\_method="semilagrangian"}), the incremental advection-based mask that confines stochastic rain generation to the vicinity of observed echoes (\texttt{mask\_method="incremental"}), and CDF-based probability matching against the observed intensity distribution (\texttt{probmatching\_method="cdf"}). All other parameters take pysteps defaults. Member~0 is used for single-member (CSI) scores and all 16 members for probabilistic (pFSS) scores.

\newpage
\section{NowcastNet evaluation details}\label{secA:NowcastNet}

Direct comparison of StormScope and NowcastNet with observations requires care:
the two systems are natively validated against different observational
fields---MRMS base reflectivity (in dBZ) on the
StormScope side, and the MRMS quantitative precipitation estimate (QPE, \unit{mm\,h^{-1}},
regridded onto the NowcastNet per-case footprint) on the
NowcastNet side. To place both forecasts in a common
scoring framework, we evaluate all threshold-based metrics
(pFSS, object detections, ensemble calibration curves, etc.) on calibrated quantile thresholds $\hat{q}_{\rm{eval}}$
rather than manually chosen thresholds.

To establish the calibrated quantiles, we select a subset of 200 storm cases within the 2024 validation set and examine the respective ground truths in dBZ and QPE against which each model would be evaluated.
From these cases, we extract the rainy-pixel population from each truth
field using a configurable floor threshold (\(\mathrm{dBZ}_{\text{thr}}\)
for reflectivity, \(\mathrm{QPE}_{\text{thr}}\) for precipitation), pool
those pixels across all cases in the calibration set, and compute the quantile thresholds for evaluation $\hat{q}_{\rm{eval}}$ from each population. The resulting $\hat{q}_{\rm{eval}}$ are then used for all downstream
statistical scoring. For example, for the two-stage NSSL object-detection procedure, we use the 90th and 95th percentiles,
\(\hat{q}_{90}\) and \(\hat{q}_{95}\), for the \emph{initialization} and \emph{maximum} thresholds, respectively. The choice of floor thresholds (\(\mathrm{dBZ}_{\text{thr}}, \mathrm{QPE}_{\text{thr}}\)) is partly arbitrary but materially
affects the rainy-population distribution, and through it the percentile
values used for detection; if the two truth pipelines do not sample
comparable storm structures, any downstream skill difference between the
two forecast systems conflates true forecast performance with calibration
mismatch.

\begin{figure}[t]
  \centering
  \includegraphics[width=0.96\linewidth]{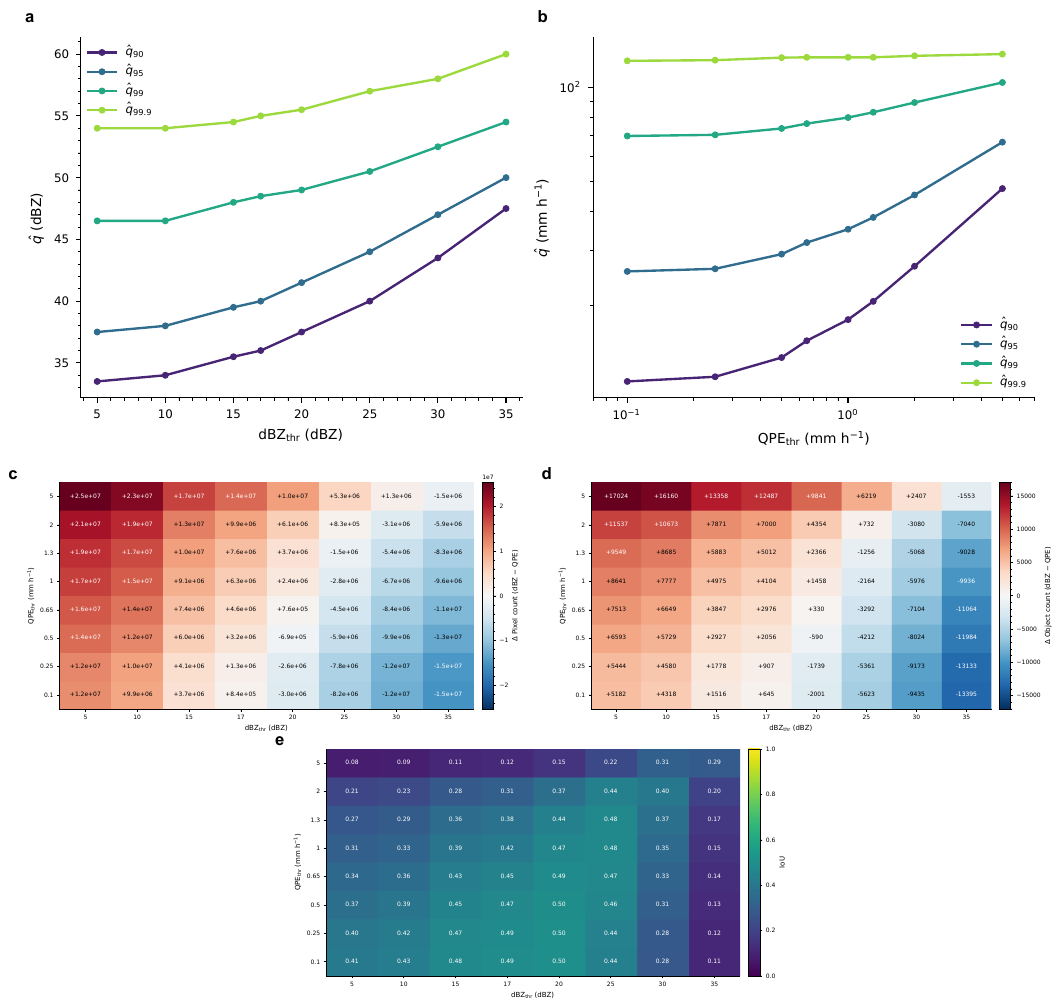}
  \caption{\textbf{NowcastNet calibration and truth-field agreement.}
    \textbf{a,b}, Calibrated quantile thresholds \(\hat{q}_{\rm{eval}}\) as a function of
    the rainy-population floor threshold for \textbf{a}, the dBZ truth
    against \(\mathrm{dBZ}_{\text{thr}}\), and \textbf{b}, the QPE truth
    against \(\mathrm{QPE}_{\text{thr}}\) on log--log axes (mirroring the
    implicit log scaling of dBZ). Lines are colored sequentially
    from lower to higher percentiles; \(\hat{q}_{90}\) and \(\hat{q}_{95}\)
    are the NSSL two-stage detection thresholds that propagate into the
    downstream object-based scoring.
    \textbf{c,d}, Agreement diagnostics between truth fields from the
    \(8\times8\) sensitivity sweep over
    \((\mathrm{dBZ}_{\text{thr}}, \mathrm{QPE}_{\text{thr}})\) on the
    2024 development subset of 200 storm cases.
    \textbf{c}, Pooled MRMS dBZ pixel count minus pooled NowcastNet QPE pixel count.
    \textbf{d}, Segmented-object count for the dBZ-based truth minus that for the
    QPE-based truth, summed over all case frames; detection uses
    \(\hat{q}_{90}\) and \(\hat{q}_{95}\) as the initialization and maximum thresholds, respectively.
    White (zero) marks pairs that yield matching counts. The composite minimum-disagreement region
    (\(\mathrm{dBZ}_{\text{thr}} \approx 17\text{--}20\),
    \(\mathrm{QPE}_{\text{thr}} \approx 0.1\,\mathrm{mm\,h^{-1}}\)) is
    visible as the near-white band in both panels.
    \textbf{e}, Mean foreground-mask IoU between the object segmentations derived
    from the dBZ- and QPE-based truth fields for each calibration pair, averaged over all
    case-frames; higher is better. The maximum IoU agreement occurs in roughly the same region
    of minimum disagreement in \textbf{c,d}.}
  \label{fig:nowcastnet-calibration}
\end{figure}

We therefore perform a sensitivity sweep over an \(8\times8\) grid in
\((\mathrm{dBZ}_{\text{thr}}, \mathrm{QPE}_{\text{thr}})\) on the 2024
validation subset of 200 storm cases. For each pair, we record the pooled
rainy-pixel populations and resulting \(\hat{q}_{90}\), \(\hat{q}_{95}\), \(\hat{q}_{99}\),
and \(\hat{q}_{99.9}\) percentile values (Fig.~\ref{fig:nowcastnet-calibration}a,b), the
total rainy-pixel and segmented-object counts
(Fig.~\ref{fig:nowcastnet-calibration}c,d), and the mean
foreground-mask IoU between the StormScope-truth and NowcastNet-truth
storm-object segmentations (Fig.~\ref{fig:nowcastnet-calibration}e). The diagnostics consistently
favor a permissive but non-trivial floor on each side: pairs with
\(\mathrm{dBZ}_{\text{thr}} \approx 17\text{--}20\) and
\(\mathrm{QPE}_{\text{thr}} \approx 0.1\text{--}0.65\,\unit{mm\,h^{-1}}\) minimize the
StormScope--NowcastNet truth disagreement in both pixel count and object
count, while keeping the mean IoU near its plateau maximum across the swept
range. We adopt \(\mathrm{dBZ}_{\text{thr}} = 20\)~dBZ and \(\mathrm{QPE}_{\text{thr}} = \qty{0.5}{\milli\metre\per\hour}\) for the production calibration in all subsequent threshold-based comparisons between NowcastNet and StormScope. The specific choice of these thresholds is guided by our analysis and by canonical definitions of ``light rain.''

\section{WoFS bias correction}\label{secA:wofs_bias_correction}

\begin{figure}
\centering
\includegraphics[width=\linewidth]{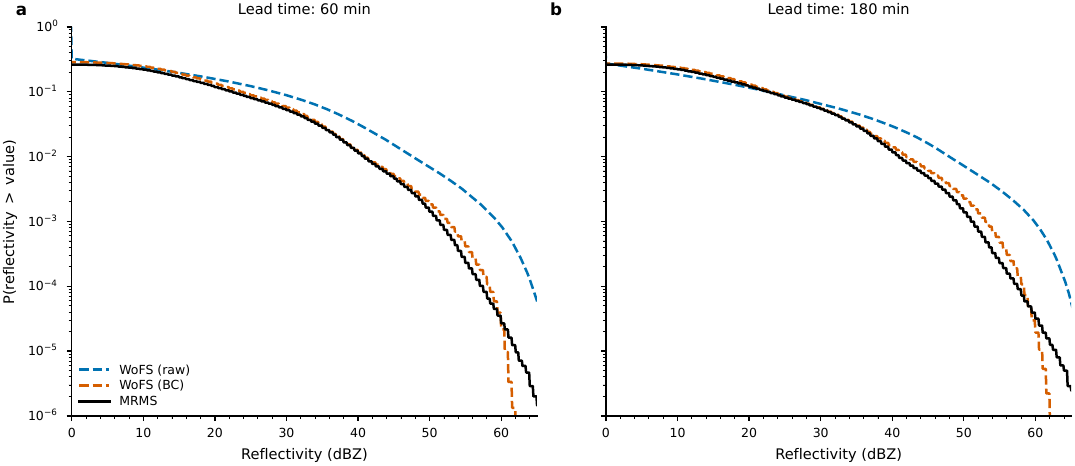}
\caption{\textbf{Warn-on-Forecast System (WoFS) reflectivity bias correction.} Each curve gives the exceedance probability---the fraction of grid pixels for which composite (column-maximum) radar reflectivity exceeds the value on the $x$-axis (in dBZ)---pooled over every pixel and every forecast at a fixed lead time. \textbf{a}, Lead time of 60~min after forecast initialization. \textbf{b}, Lead time of 180~min. Three distributions are compared in each panel: the uncorrected WoFS ensemble forecast (WoFS raw, blue dashed); the same forecast after applying an empirical per-lead quantile-mapping correction that remaps each WoFS reflectivity to the radar reflectivity with the same percentile in the training distribution (WoFS BC, vermilion dashed); and the Multi-Radar/Multi-Sensor (MRMS) ground-based radar mosaic that serves as the verification target (black solid). The correction is fit entirely on 200 independent 2024 forecasts and is never exposed to the 2025 forecasts shown here. One ensemble member per forecast is pooled across 100 randomly selected 2025 forecasts to construct each distribution. Without correction, WoFS overproduces spurious heavy reflectivity. Operational meteorologists typically account for WoFS biases either explicitly through bias correction or implicitly through familiarity with the model outputs over an extended period.}
\label{fig:wofs-bias-correction}
\end{figure}

Uncorrected WoFS composite reflectivity overproduces heavy reflectivity relative to MRMS verification (Fig.~\ref{fig:wofs-bias-correction}). We therefore apply an empirical per-lead-time quantile mapping that remaps each WoFS reflectivity value to the MRMS reflectivity at the same percentile of the training distribution. The correction is fit on 200 independent 2024 forecasts and is never exposed to the 2025 evaluation set. All WoFS results labeled ``BC'' use this correction.

\clearpage

\section{Domain parallelization and scaling}
\label{secA:domain_parallel}

Geostationary satellite data are high resolution and span large spatial domains, so we developed the StormScope framework with these requirements in mind. Previous work has adopted the MultiDiffusion~\citep{bar2023multidiffusion,brenowitz2025climate} approach, which runs the denoising model on smaller local patches during training to save memory and computation; during sampling, the patches are stitched together via boundary overlaps at each step. Although effective, this approach induces a tradeoff between efficiency and sample quality or coherence (see the discussion in Ref.~\citep{brenowitz2025climate}) and limits the spatial context available to the model, depending on the local patch size. To sidestep these concerns, we built a model and training procedure that can easily leverage domain parallelism to scale both training and inference directly to larger domains, including full-disk forecasts. We demonstrate this capability by training a \qty{3}{\kilo\metre} satellite nowcast model over the CONUS domain, spanning \(1792 \times 1024\) grid points (\(\approx 1.8\) million pixels).

At this scale, the memory required for standard model training exceeds the capacity of typical data-center GPUs ($\sim$80~GB), even when using a local batch size of 1, necessitating some form of model parallelism to split a model instance across multiple GPUs. Because the input resolution, and thus the number of tokens, for such a model is so large, the bulk of the memory is consumed by intermediate activations, so parameter-sharding techniques such as Fully Sharded Data Parallel (FSDP) alone cannot solve the problem. Thus, training the \qty{3}{\kilo\metre} model requires model parallelism that specifically targets the per-GPU memory consumed by model activations, as employed in other large-scale AI weather models~\citep{lang2024aifs, bonev2025fourcastnet}. Specifically, we use domain (spatial) parallelism on a two-dimensional device mesh with the \texttt{ShardTensor} API from NVIDIA PhysicsNeMo, which requires minimal changes to the model code.

The two-dimensional spatial domain is sharded into upper and lower portions, which are processed by a pair of GPUs comprising one model instance. The GPUs are organized such that those within a domain-parallel group exchange information as necessary during forward and backward passes, then reduce gradients across the data-parallel groups after the backward pass, as in standard data-parallel training. Because the underlying architecture is a Diffusion Transformer (DiT) with Neighborhood Attention (NA), most internal operations in the model (e.g., tokenization and elementwise MLPs) are embarrassingly parallel across the domain-parallel axis and require no communication during the forward pass. The only exceptions are operations that aggregate or mix features across the token sequence (spatial domain), which in our case are the two-dimensional Neighborhood Attention kernels. To ensure that the output is correctly computed across shard interfaces, halo exchanges propagate contextual information for border tokens. By using NA windows, we reduce the volume of communication needed because only the halo, rather than the full token sequence, must be exchanged. The width of each halo corresponds to the kernel size of the NA operator. The loss is computed locally on each GPU and subsequently reduced across the domain group so that every device possesses the aggregated loss for the full spatial domain before backpropagation.

Because domain parallelism for two-dimensional NA is supported in PhysicsNeMo \texttt{ShardTensor}, all sharding and communication operations are handled automatically, and the only code changes necessary are in the training loop to initialize the device mesh and shard the input data. A significant advantage of this approach is its scalability and ease of use, as the model can be trained on larger spatial domains or higher-resolution data by simply increasing the domain-parallel group size without modifying the model architecture. To assess the scaling efficiency of our implementation, we evaluate the ratio of iteration throughput for the domain-parallel implementation to that of the standard data-parallel-only baseline, measured using an identical number of GPUs. Using the described configuration, with a domain-parallel size of 2 and a data-parallel size of 4, we achieve a scaling efficiency of approximately 87.5\,\% for the CONUS-wide \qty{3}{\kilo\metre} satellite model, demonstrating that domain parallelism enables training over much larger domains while retaining most of the computational throughput.

\section{Extended calibration details}\label{secA:Calibration}

\begin{figure}[H]
    \centering
    \includegraphics[width=0.9\linewidth]{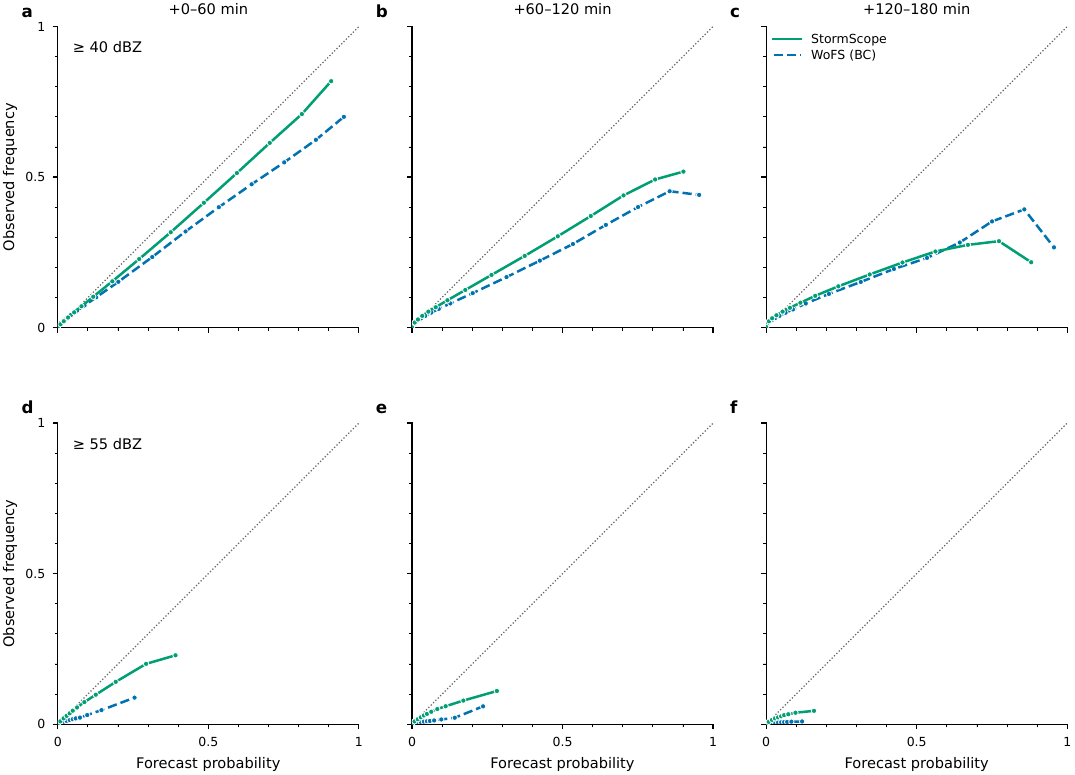}
    \caption{Lead-time-stratified reliability diagrams comparing StormScope (green solid) and WoFS (blue dashed) ensembles at the 40-dBZ composite-reflectivity threshold (\textbf{a--c}) and the 55-dBZ threshold (\textbf{d--f}), stratified by forecast lead times of 0--60~min (\textbf{a}, \textbf{d}), 60--120~min (\textbf{b}, \textbf{e}), and 120--180~min (\textbf{c}, \textbf{f}). Gray dotted diagonals indicate perfect reliability.}
    \label{fig:calibration_wofs_ss}
\end{figure}

\begin{figure}[H]
    \centering
    \includegraphics[width=0.9\linewidth]{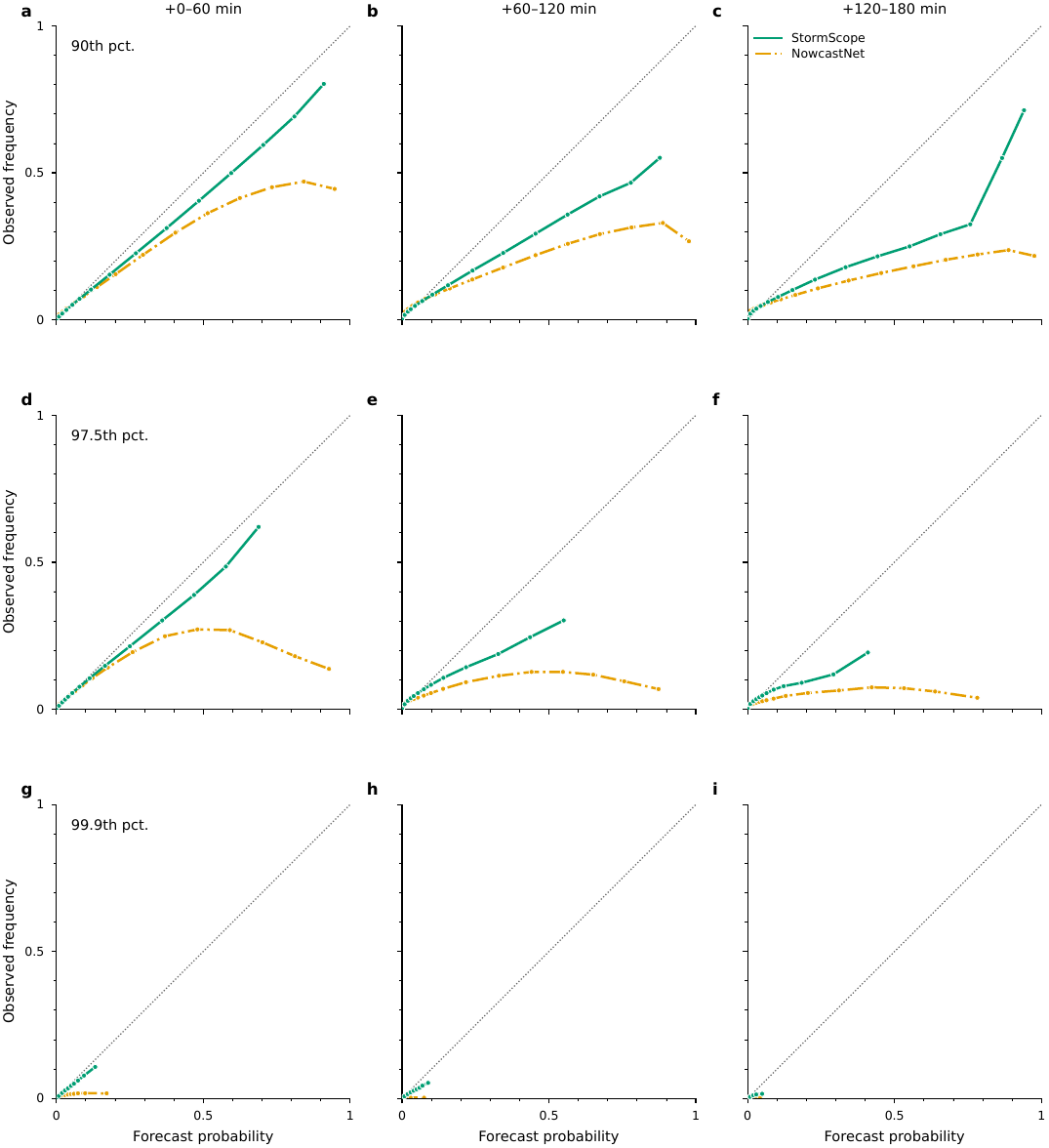}
    \caption{Lead-time-stratified reliability diagrams comparing StormScope (green solid) and NowcastNet (orange dash-dot) ensembles at the 90th-percentile threshold (\textbf{a--c}), 97.5th-percentile threshold (\textbf{d--f}), and 99.9th-percentile threshold (\textbf{g--i}), stratified by forecast lead times of 0--60~min (\textbf{a}, \textbf{d}, \textbf{g}), 60--120~min (\textbf{b}, \textbf{e}, \textbf{h}), and 120--180~min (\textbf{c}, \textbf{f}, \textbf{i}). Gray dotted diagonals indicate perfect reliability.}
    \label{fig:calibration_nn_ss}
\end{figure}

Figures~\ref{fig:calibration_wofs_ss} and~\ref{fig:calibration_nn_ss} show the reliability diagrams of Fig.~\ref{fig:combined_skill_figure}, stratified by forecast lead times of 0--60~min, 60--120~min, and 120--180~min. Figure~\ref{fig:calibration_wofs_ss} shows that StormScope is better calibrated, lying closer to the 1:1 diagonal at both thresholds during the first 2~h. Both models are underdispersive at longer lead times. Figure~\ref{fig:calibration_nn_ss} shows that NowcastNet ensemble forecasts are dramatically overconfident and underdispersive at all thresholds and lead times, while StormScope provides better-calibrated ensemble guidance.

\clearpage

\section*{Extended Data tables}
\renewcommand{\tablename}{Extended Data Table}
\renewcommand{\thetable}{\arabic{table}}
\setcounter{table}{0}

\begin{table}[!htbp]
  \centering
  \caption{Skill of bias-corrected WoFS and StormScope on the 2025
  WoFS-initialized evaluation set ($n = 929$ matched cases) as a
  function of forecast lead time. Object-based Critical Success Index (CSI)
  uses the NSSL 40/45~dBZ two-threshold detector and total-interest matching.
  Neighborhood-grid probability of detection (POD), success ratio (SR) and CSI
  are from the 33~km both-field, any-in-window grid-point contingency (member 0,
  counts pooled over all cases at each lead) at the 40- and 55-dBZ thresholds.
  Probabilistic Fractions Skill Score (pFSS) is reported at the same thresholds
  with a fixed 33~km neighborhood (11 pixels on the 3~km verification grid).
  Bold marks the better model at each lead. See Extended Data
  Fig.~\ref{fig:combined-2025} and Methods Secs.~\ref{sec:object-based-eval}
  and~\ref{sec:pfss} for details.}
  \label{tab:wofs_vs_stormscope_skill}
  \begin{tabular}{llcccc}
    \toprule
    & & \multicolumn{4}{c}{Lead time (min)} \\
    \cmidrule(lr){3-6}
    Metric & Model & 30 & 60 & 120 & 180 \\
    \midrule
    \multirow{2}{*}{Object CSI}
      & WoFS (BC)  & 0.35 & 0.29 & 0.26 & \textbf{0.24} \\
      & StormScope & \textbf{0.68} & \textbf{0.53} & \textbf{0.33} & 0.22 \\
    \midrule
    \multirow{2}{*}{Nbd. grid POD (40~dBZ, 33~km)}
      & WoFS (BC)  & 0.47 & 0.40 & 0.34 & 0.33 \\
      & StormScope & \textbf{0.75} & \textbf{0.64} & \textbf{0.50} & \textbf{0.40} \\
    \multirow{2}{*}{Nbd. grid SR (40~dBZ, 33~km)}
      & WoFS (BC)  & 0.55 & 0.48 & 0.45 & 0.44 \\
      & StormScope & \textbf{0.75} & \textbf{0.65} & \textbf{0.53} & \textbf{0.46} \\
    \multirow{2}{*}{Nbd. grid CSI (40~dBZ, 33~km)}
      & WoFS (BC)  & 0.34 & 0.28 & 0.24 & 0.23 \\
      & StormScope & \textbf{0.60} & \textbf{0.48} & \textbf{0.35} & \textbf{0.27} \\
    \midrule
    \multirow{2}{*}{Nbd. grid POD (55~dBZ, 33~km)}
      & WoFS (BC)  & 0.16 & 0.14 & \textbf{0.10} & \textbf{0.09} \\
      & StormScope & \textbf{0.44} & \textbf{0.26} & 0.09 & 0.04 \\
    \multirow{2}{*}{Nbd. grid SR (55~dBZ, 33~km)}
      & WoFS (BC)  & 0.19 & 0.16 & 0.12 & 0.10 \\
      & StormScope & \textbf{0.43} & \textbf{0.33} & \textbf{0.20} & \textbf{0.13} \\
    \multirow{2}{*}{Nbd. grid CSI (55~dBZ, 33~km)}
      & WoFS (BC)  & 0.09 & 0.08 & 0.06 & \textbf{0.05} \\
      & StormScope & \textbf{0.27} & \textbf{0.17} & \textbf{0.07} & 0.03 \\
    \midrule
    \multirow{2}{*}{pFSS (40~dBZ, 33~km)}
      & WoFS (BC)  & 0.63 & 0.53 & 0.47 & 0.44 \\
      & StormScope & \textbf{0.91} & \textbf{0.79} & \textbf{0.59} & \textbf{0.45} \\
    \multirow{2}{*}{pFSS (55~dBZ, 33~km)}
      & WoFS (BC)  & 0.26 & 0.22 & 0.14 & \textbf{0.11} \\
      & StormScope & \textbf{0.72} & \textbf{0.50} & \textbf{0.20} & 0.06 \\
    \bottomrule
  \end{tabular}
\end{table}

\clearpage

\end{appendices}

\end{document}